\documentclass[preprint,aps,prb]{revtex4-1}
\usepackage{graphics}
\usepackage{amssymb}
\usepackage{upgreek}
\usepackage[tight,TABTOPCAP]{subfigure}
\usepackage{subfloat}
\usepackage{float}
\usepackage{graphicx}
\usepackage{dcolumn}
\usepackage{bm}
\usepackage{setspace}
\usepackage{wrapfig}
\usepackage[intlimits]{empheq}
\usepackage{color}
\usepackage{relsize}

\makeatletter

\newcommand{\vect}[1]{{\boldsymbol{#1}}}

\begin{document}

\title{Distinguishing Coulomb and electron-phonon interactions for massless Dirac fermions.}
\author{J. P. F. LeBlanc$^{1,2}$}
\author{Jungseek Hwang$^{3}$}
\email{jungseek@skku.edu; Corresponding author}
\author{J. P. Carbotte$^{4,5}$}
\affiliation{$^1$Department of Physics, University of Guelph,
Guelph, Ontario N1G 2W1 Canada}
\affiliation{$^2$Guelph-Waterloo Physics Institute, University of Guelph, Guelph, Ontario N1G 2W1 Canada}
\affiliation{$^3$Department of Physics, Sungkyunkwan University, Suwon, Gyeonggi-do
440-746, Republic of Korea}
\affiliation{$^4$Department of Physics and Astronomy, McMaster
University, Hamilton, Ontario L8S 4M1 Canada}
\affiliation{$^5$The Canadian Institute for Advanced Research, Toronto, ON M5G 1Z8 Canada}

\date{\today}
\begin{abstract}
While many physical properties of graphene can be understood qualitatively on the basis of bare Dirac bands, there is specific evidence that electron-electron (EE) and electron-phonon (EP) interactions can also play an important role.  We discuss strategies for extracting separate images of the EE and EP interactions as they present themselves in the electron spectral density and related self-energies.
While for momentum, $k$, equal to its Fermi value, $k_F$, a composite structure is obtained which can be difficult to separate into its two constituent parts, at smaller values of $k$ the spectral function shows distinct incoherent sidebands on the left and right of the main quasiparticle line.  These image respectively the EE and EP interactions, each being most prominent in its own energy window.  We employ a maximum entropy inversion technique on the self energy to reveal the electron-phonon spectral density separate from the excitation spectrum due to coulomb correlations.  Our calculations show that this technique can provide important new insights into inelastic scattering processes in graphene.
\end{abstract}
%
\maketitle

\section{Introduction}
Graphene consists of a single layer of carbon atoms arranged on a hexagonal honeycomb crystal lattice which has two atoms per unit cell and two energy bands in the Brillouin zone.  The charge carriers exhibit unusual dynamics which are governed by the Dirac equation for massless Fermions, now well documented in several reviews.\cite{geim:2007,neto:2009,gusynin:2007:ijmp,abergel:2010,orlita:2010}
At low energies the electronic dispersions are linear in energy and the tip of the two cones associated with valence and conduction bands respectively meet at a single Dirac point where the electronic density of states (DOS) vanishes.  While a bare band picture provides a good description of many of the observed properties of the charge dynamics of graphene, signatures of many body corrections provided by the electron-phonon as well as electron-electron interactions have also been seen. \cite{geim:2007,neto:2009,gusynin:2007:ijmp,abergel:2010,orlita:2010}
For example, kinks appear in the dressed quasiparticle energies measured by angular resolved photoemission spectroscopy (ARPES).\cite{bostwick:2007, bianchi:2010, zhou:2008}
These structures can, in part, represent coupling of the electronic system to bosonic modes.\cite{carbotte:2011,schachinger:2000, schachinger:2003}  The mode involved could be a phonon\cite{bostwick:2007, bianchi:2010} but in some systems such as, for example, in the high critical-temperature superconducting cuprates the boson mode may be spin fluctuations that have their origin in the strongly correlated nature of these materials.\cite{carbotte:2011}  In graphene, structures corresponding to coupling to a variety of bosons have been seen in scanning tunnelling spectroscopy (STS).\cite{li:2009,miller:2009,brar:2010, pound:2011,nicol:2009}
This is not expected in conventional metals for which electronic density of states, $N(\omega)$, around the Fermi level is constant on the energy scale set by the boson.  However, if instead $N(\omega)$ varies on a scale comparable to the boson energy structure, as is the case in graphene, then this structure should be seen.\cite{mitrovic:1983,mitrovic:1983:2}
There also exists evidence for modifications of the bare dispersions due to formation of plasmarons which are due to electron-electron interactions.  Experimental ARPES\cite{bostwick:2010} spectra show that the Dirac point, associated with the point of coincidence of the valence and conduction bands is split into two and an extended plasmaron region is observed between these two points.

In metallic systems, renormalizations due to electron-electron or electron-phonon interactions can have profound effects on the physical properties of their normal as well as superconducting states.\cite{carbotte:1995,nicol:1991,nicol:1991:2,schachinger:1997}
These interactions provide inelastic scattering\cite{carbotte:1995,nicol:1991} and lead to strong coupling corrections to BCS results.\cite{carbotte:1986,marsiglio:1992,mitrovic:1980}
There are also other complications that can affect properties, such as the presence of a van-Hove singularity\cite{schachinger:1990,arberg:1993} and scattering anisotropies\cite{branch:1995,leung:1976,odonovan:1995} but these are not expected to be important in discussions of the low energy properties of graphene due to its unique band structure near the Fermi level.

In this paper we consider the combined effect of both electron-electron interactions (EEI) and electron-phonon interactions (EPI) on properties of doped graphene with a view at understanding how they present themselves as boson structure and with a particular emphasis on the possibility of distinguishing these two interactions from each other.  The EPI is not expected to be large, having a mass enhancement factor, $\lambda$, which in graphene corresponds to a constant reduction of the Fermi velocity at the Fermi surface by a factor of $(1+\lambda)$.  In the literature $\lambda$ varies considerably from less than 0.1 \cite{park:2007,park:2008,park:2009} in some calculations and in the interpretation\cite{pound:2011} of the experimental data of Ref.~\onlinecite{miller:2009} to larger values in scanning tunneling micsrocopy\cite{li:2009, nicol:2009} and to more than 0.3 in other experimental estimates.\cite{bostwick:2007}  We hope that the work presented here can help in providing guidance to more reliable experimental estimates of the size of $\lambda$.

While in the end the $\lambda$ in graphene may turn out to be small, optical conductivity measurements have observed Holstein type side bands in graphene. These provide considerable absorption in the region between the Drude peak,\cite{li:2008,wang:2008,mak:2008,nair:2008} centered at zero photon energy, and the sharp rise towards the universal conductivity at twice the chemical potential.  The chemical potential can be made large by doping or charging in a field effect configuration.  In this way one can have a large window of photon energy where the bare band model would predict essentially zero conductivity, while present experiments find a conductivity that is almost one third of the universal value $\sigma_0$.  This region, of course, can be filled by correlation effects such as the EPI for which there have been several estimates for the conductivity.\cite{peres:2008,carbotte:2010}
These estimates largely agree with each other and conclude that the EPI on its own, is unlikely to account for the experimental observation.  Additional filling of the Pauli blocked region of bare bands can come from impurity effects and from the EEI.\cite{grushin:2009}  Note that such effects are quite distinct from bilayer signatures; a system that has also been extensively studied.\cite{nicol:2008,kuzmenko:2009,henriksen:2008}

As we have already mentioned, electron-electron effects have been observed in ARPES experiments in the form of a plasmaron band and the splitting of the Dirac point into two.\cite{bostwick:2010}  There the data is in good agreement with the G$_0$W-RPA approximation.  This calculation involves a dynamically screened potential based on random phase screening.\cite{wunsch:2006,polini:2008,hwang:2008, sensarma:2011,barlas:2007,ehhwang:2007,leblanc:2011} The resulting effective electron-electron interaction depends inversely on the size of the average substrate dielectric constant, $\epsilon$, on either side of the graphene layer.  Consequently the size of the plasmaron structure in the ARPES curves\cite{bostwick:2010} depends on $\epsilon$ as does the equivalent structure seen in the corresponding density of states.\cite{leblanc:2011,principi:2011}  These structures are quite distinct from phonon structure as we will study in this paper. In Sec.~\ref{sec:2} we summarize, from the existing literature, the formulas needed to calculate the self energies we wish to consider; the EEI within a random phase approximation and the EPI for a general form of the electron-boson spectral density.  In actual numerical calculations we use a phonon model consisting of one or two truncated lorentzian peaks.  Based on these self energies we present results for the charge carrier spectral density, $A(k,\omega)$, for momentum, $k$, and energy, $\omega$, for EEI and EPI alone as well as combined and we discuss how each interaction manifests.  In Sec.~\ref{sec:inversion} we introduce a maximum entropy technique which allows us to obtain an effective electron-boson spectral density, $\alpha^2F(\omega)$, given a self energy as a starting point.  The $\alpha^2F(\omega)$ function will reveal clearly the underlying spectrum of bosonic excitations involved in the quasiparticle scattering.  This is an exact procedure for the EPI but as we will see it is also useful for the case of EEI for which the concept of a boson exchange mechanism is only approximate.  Our conclusions are given in Sec.~\ref{sec:conclusions}.

\section{Self energies for Electron-Electron (EE) and Electron-Phonon (EP) Interactions}\label{sec:2}

\begin{figure}
\vspace*{-0.5cm}%
\centerline{\includegraphics[width=3.5 in]{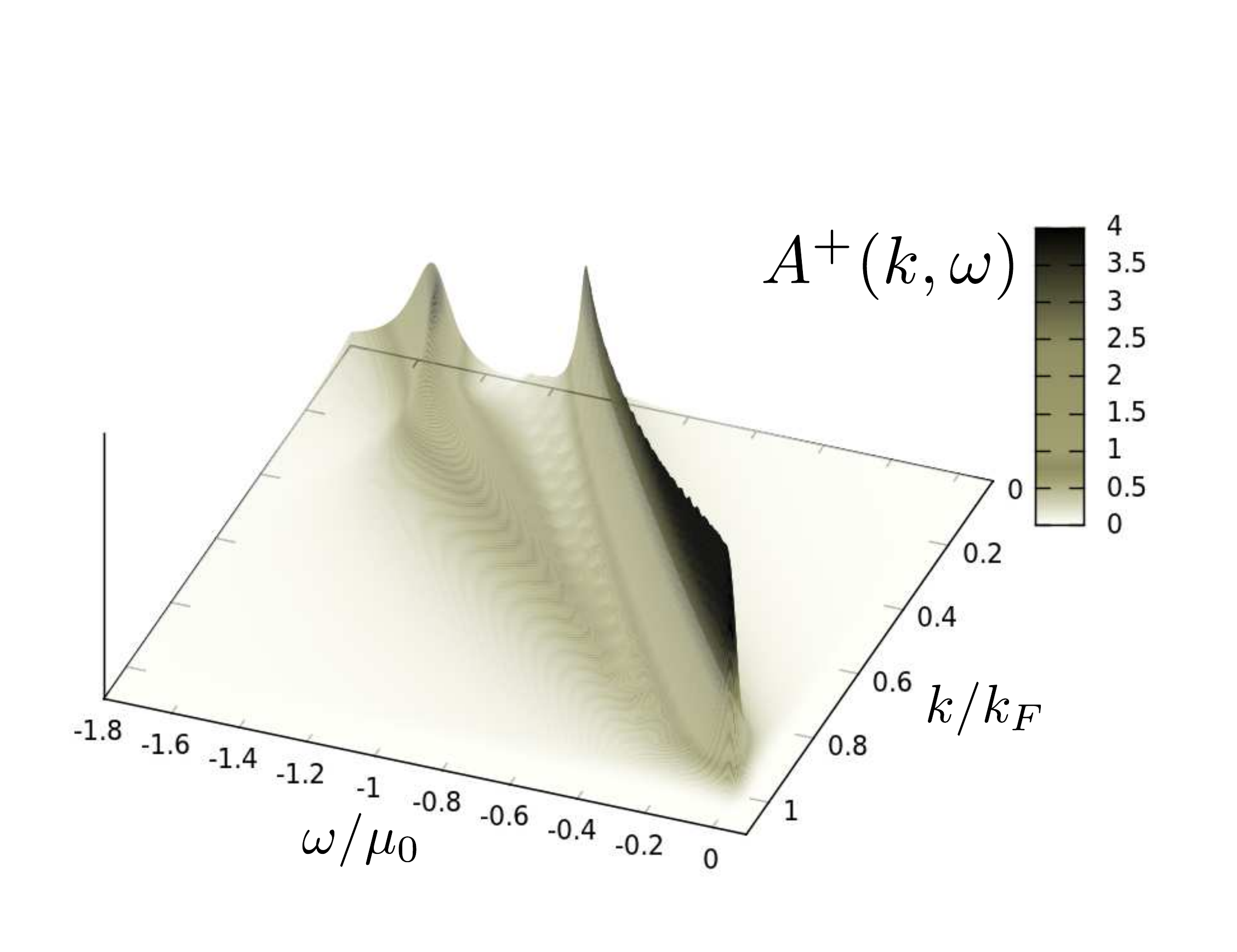}}
\vspace*{-0.5cm}%
  \caption{\label{fig:colourplot}(Color online)  Plot of spectral function, $A^+(k,\omega)$ for the $s=1$ band including the EEI.  The primary band has a plasmaron sideband which approaches the main band at the Fermi level at $k=k_F$. }
\end{figure}

\begin{figure}
\vspace*{-0.5cm}%
\centerline{\includegraphics[width=4.0 in]{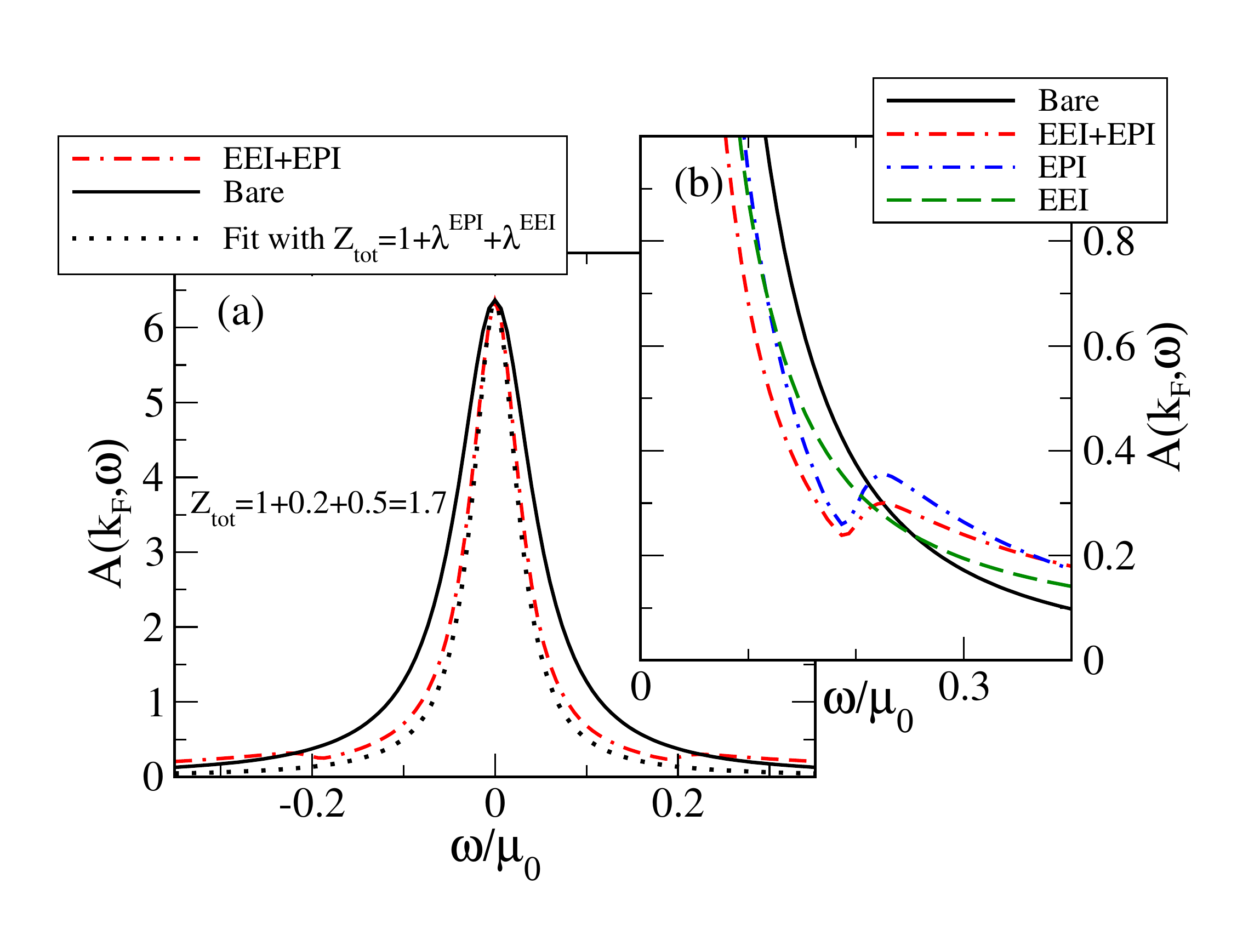}}
\vspace*{-0.5cm}%
  \caption{(Color online) (a)The electronic spectral density, $A(k_F,\omega)$, as a function of $\omega$ normalized to the chemical potential $\mu_0=1$~eV.  The case including EEI and EPI is shown in the dotted-double-dashed red line.  Also shown for comparison is the bare band case (solid black curve) with a residual scattering rate $\Gamma/\mu_0=0.05$.    The black dotted curve is a Lorentzian which is derived analytically in the text [see Eq.~(\ref{eqn:akwz})]. (b) The electronic spectral function $A(k_F,\omega)$ as a function of $\omega/\mu_0$ plotted for several cases for comparison.  Emphasis is on the wing region out from the main quasiparticle peak centered on $\omega=0$.\label{fig:2} }
\end{figure}

Details associated with the calculation of the self energy associated with electron-electron interactions (EEI) in graphene are well documented in the literature  and will not be repeated here.\cite{wunsch:2006,polini:2008, hwang:2008,sensarma:2011, barlas:2007, ehhwang:2007, sensarma:2011, leblanc:2011, principi:2011}  We follow the notation established in Ref.~\onlinecite{leblanc:2011} where their Eq.~(A8) and (A9) provide the necessary expressions for the quasiparticle self energy, $\Sigma_s(k,\omega)$, where $s=\pm$ is the label for the conduction and valence band respectively.  In dimensionless units $\bar{\omega}=\omega/\mu_0$, $\bar{k}=k/k_F$ and $\bar{\Sigma}_s(\bar{k},\bar{\omega})\equiv \Sigma_s(k,\omega)/\mu_0$ where $\mu_0$ is the chemical potential.  The relevant equations are
\begin{widetext}
\begin{align}\label{eqn:res}
\bar{\Sigma}_s^{RES}(\bar{k},\bar{\omega})=\sum_{s^\prime=\pm 1}\int_{0}^\infty \int_{0}^{2\pi} &\frac{d\bar{q}d\theta_{\boldsymbol{kq}}}{2\pi}\frac{\alpha}{g}\upvarepsilon^{-1}(\bar{q},\bar{\omega}-\bar{\epsilon}_{\vect{k}+\vect{q}}^{s^\prime})F_{ss^\prime}(\theta_{\vect{kk}^\prime})\times\nonumber \\&[\Theta(\bar{\omega}-\bar{\epsilon}_{\vect{k}+\vect{q}}^{s^\prime})-\Theta(-\bar{\epsilon}_{\vect{k}+\vect{q}}^{s^\prime})],
\end{align}
and
\begin{align}\label{eqn:line}
\bar{\Sigma}_s^{line}(\bar{k},\bar{\omega})=&-\sum_{s^\prime=\pm 1}\int_{0}^\infty \int_{0}^{2\pi}\frac{d\bar{q} d\theta_{\boldsymbol{kq}}}{2\pi}\frac{\alpha}{g}F_{ss^\prime}(\theta_{\vect{kk}^\prime})\times \nonumber\\ &\int_{-\infty}^{\infty}\frac{d\bar{\Omega}}{2\pi}\upvarepsilon^{-1}(\bar{q},\imath \bar{\Omega})\Bigg[\frac{\bar{\omega}-\bar{\epsilon}_{\vect{k}+\vect{q}}^{s^\prime}}{\bar{\Omega}^2+(\bar{\epsilon}_{\vect{k}+\vect{q}}^{s^\prime}-\bar{\omega})^2}-\frac{\imath \bar{\Omega}}{\bar{\Omega}^2+(\bar{\epsilon}_{\vect{k}+\vect{q}}^{s^\prime}-\bar{\omega})^2}\Bigg],
\end{align}
\end{widetext}
where the total self energy for band $s$ due to the electron-electron interaction is
\begin{equation}\label{eqn:eei}
\bar{\Sigma}_s^{EEI}(\bar{k},\bar{\omega}) = \bar{\Sigma}_s^{RES}(\bar{k},\bar{\omega}) +\bar{\Sigma}_s^{line}(\bar{k},\bar{\omega}).
\end{equation}
In these equations, which are derived for the G$_0$W-RPA approximation,  $\upvarepsilon^{-1}$ is the inverse dielectric function, $\alpha=\frac{ge^2}{\epsilon_0 v_F}$ controls the overall strength of Coulomb potential, with a degeneracy factor $g=4$ for graphene two valleys and two spins and $\epsilon_0$ is the bare dielectric constant.  Due to the 2-dimensional nature of graphene, its bare dielectric constant is the average of values of the materials above and below the graphene sheet.  Thus, this parameter can be varied over a significant range by changing the substrate material.\cite{walter:2011}
The band energies in the Dirac cone approximation for general $\vect{k}$ and $\vect{q}$ in a dimensionless form are
\begin{align}
\bar{\epsilon}_{\vect{k}+\vect{q}}^{s^\prime}=\frac{\epsilon_{\vect{k}+\vect{q}}^{s^\prime}}{\mu_0}=s^\prime\sqrt{\bar{k}^2+\bar{q}^2+2\bar{k}\bar{q}\cos\theta_{\vect{k}\vect{q}}}-1
\end{align}
\normalsize
and the band overlaps are given by
\begin{equation}
F_{s s^\prime}(\theta_{\vect{k}\vect{k}^\prime})=\frac{1}{2}[1+\cos(\theta_{\vect{k}\vect{k}^\prime})s s^\prime],
\end{equation}
where $\theta_{\boldsymbol{kk^\prime}}$ is the angle between the vector $\vect{k}$ and the scattering vector $\vect{k^\prime}$ and is related to the integration variable $\theta_{\boldsymbol{kq}}$, the angle between vectors $\boldsymbol{k}$ and $\boldsymbol{q}$.

As can be seen from the normalization, the EEI self energy scales with $\mu_0$.  Once we add a contribution from the interaction with phonons, an energy scale is introduced into the problem, namely the phonon energy, $\omega_E$, and the value of the chemical potential $\mu_0$ needs to be specified relative to $\omega_E$.

Detailed calculations of the electronic self energies in graphene due to electron-phonon coupling have been done in density function theory.\cite{park:2007,park:2008,park:2009}  An observation, important for the present work, made in Ref.~\onlinecite{park:2007}, is that the results of such complex computations show little dependence on the direction and magnitude of the electron momentum, $\boldsymbol{k}$, and can be modeled in a first approximation through coupling to a single phonon mode at energy $\omega_E=200$~meV.  Here we adopt this model but will also allow for coupling  to a group of phonons (still assumed independent of momentum) rather than a single mode.  This can be accomplished by the introduction of an electron-boson spectral density, $\alpha^2F(\omega)$, which is independent of the Dirac fermion momentum, $\boldsymbol{k}$.  With such a simplified model,
the self energy coming from the electron-phonon interaction (EPI) takes the form\cite{mitrovic:1983,nicol:2009}
\begin{widetext}
\begin{align}
\Sigma^{EPI}(k,\omega)=\int_{0}^{\infty} \alpha^2 F(\nu)d\nu \int_{-\infty}^{\infty} d\omega^\prime \frac{N(\omega^\prime)}{N_0}\left[ \frac{n(\nu)+f(-\omega^\prime)}{\omega-\nu-\omega^\prime+i0^+}  +\frac{n(\nu)+f(\omega^\prime)}{\omega+\nu-\omega^\prime+i0^+}\right].\label{eqn:episigma}
\end{align}
\end{widetext}
Here $n(\nu)$ and $f(\nu)$ are the Bose and Fermi distribution functions which at zero temperature reduce respectively to $0$ and a Heaviside function, $\Theta$, where we have arranged that the Fermi energy falls at $\omega^\prime=0$, ie. $\frac{N(\omega^\prime=0)}{N_0}=1$.  In Eq.~(\ref{eqn:episigma}), the DOS is set equal to one at $\omega^\prime=0$.  In this work we will model the electron-phonon spectral density, $\alpha^2F(\omega)$ with either one or two truncated Lorentzians given by
\begin{equation}
\alpha^2 F(\nu)=\frac{A^\prime}{\pi}\left[ \frac{\delta}{(\nu-\omega_E)^2+\delta^2}  - \frac{\delta}{\delta_c^2+\delta^2}             \right]\Theta\left(\delta_c -|\omega_E-\nu|      \right).
\end{equation}
This creates a phonon intensity distributed in a truncated Lorentzian shape centered about the value $\omega_E$.  Though arbitrarily chosen in this work, the values of $\delta$ and $\delta_c$ are taken to be 15~meV and 30~meV respectively for each Lorentzian used.  This form is in lieu of a single $\delta$ function phonon mode and provides widths to the $\alpha^2F(\omega)$ that will allow us to more rigorously check the results of the maximum entropy inversion that will follow in Section \ref{sec:inversion}.
We refer to the mass enhancement factor, $\lambda=2\int\limits_0^\infty \frac{\alpha^2F(\nu)}{\nu}d\nu$, that sets the scale on $\alpha^2 F(\nu)$ which is dimensionless, and $A^\prime$ is varied to get the desired value of $\lambda$.

The electron spectral density $A(k,\omega)$ for momentum, $k$, and energy, $\omega$, determines the single particle properties of graphene.  Including interactions we have
\begin{widetext}
\begin{equation}
A(k,\omega)=\sum_{s=\pm} A^s(k,\omega)\equiv \sum_{s=\pm}\frac{1}{\pi}\frac{-{\rm Im}\Sigma_s(k,\omega)}{\left[\omega-\epsilon_k^s-{\rm Re}\Sigma_s(k,\omega)\right]^2+\left[ {\rm Im}\Sigma_s(k,\omega)  \right]^2}\label{eqn:akw}
\end{equation}
\end{widetext}
where $s=\pm$ are the conduction and valence bands respectively.  Here, $\epsilon_k^s=s v_F |k|-\mu$ where the shift in chemical potential, ${\rm Re}\Sigma(k_F,\omega=0)$, has been absorbed into $\epsilon_k^s$ such that $\mu= \mu_0 +{\rm Re}\Sigma(k_F,\omega=0)$. In our case the total self energy in each band is the sum of the EEI and the EPI.  To illustrate our results for the EEI alone we show in Fig.~\ref{fig:colourplot} a three dimensional plot of the spectral density (vertical axis) as a function of normalized energy, $\omega/\mu_0$, and momentum, $k/k_F$.  In these normalized units this single plot represents all doping levels.  We have in mind chemical potentials of order several hundred meV to an electron volt.  Around $k=k_F$, we see a single quasiparticle peak with finite width and a small background.  As momentum is reduced towards $k=0$, a prominent second peak forms at energies below that of the main quasiparticle peak.
This structure represents the formation of what is called a plasmaron band and is associated with the formation of reasonably long lived collective modes (plasmarons) between charge oscillations (plasmons) and the particle-hole continuum which are a direct result of electron-electron interactions.
Cuts in the spectral function for constant energy, $\omega$, are referred to as the momentum distribution curves (MDC) while cuts for constant momentum, $k$, give the energy distribution curves (EDC) which can have a complicated, non-Lorentzian dependence on $\omega$ while the MDC are much closer to pure Lorentzian.
This arises because the self energy at fixed momentum can have a dependence on energy.

For the case of $k=k_F$, if we expand the self energy in powers of $\omega$ we find that
\begin{equation}
\Sigma(k_F,\omega)\cong \Sigma(k_F,0) + \omega(\lambda^{EEI}+\lambda^{EPI})
\end{equation}
for small $\omega$ where
\begin{equation}
\lambda^{EEI}=-\left.\frac{\partial \Sigma^{EEI}(k_F,\omega)}{\partial \omega}\right|_{\omega=0}
\end{equation}
and
\begin{equation}
\lambda^{EPI}=-\left.\frac{\partial \Sigma^{EPI}(k_F,\omega)}{\partial \omega}\right|_{\omega=0}.
\end{equation}
The first constant term renormalizes the chemical potential from its dressed to bare value as we already accounted for and plays no role beyond this.  Substitution of these approximate results into Eq.~(\ref{eqn:akw}) gives a Lorentzian form for the positive branch as
\begin{equation}\label{eqn:akwz}
A^+(k_F,\omega)=\frac{1}{\pi Z^{tot}}\frac{\Gamma/Z^{tot}}{\omega^2+(\Gamma/Z^{tot})^2}
\end{equation}
where $Z^{tot}=1+\lambda^{EEI}+\lambda^{EPI}$.  The case shown in Fig.~\ref{fig:2}(a) as the black dotted curve is the numerical evaluation of Eq.~(\ref{eqn:akwz}), shown for $\lambda^{EEI}=0.5$ and $\lambda^{EPI}=0.2$, while the full $A(k,\omega)$ is shown in the double-dash-dotted red curve.  These two curves follow each other reasonably well for small $\omega$.  Thus, at small $\omega$, the dressed quasiparticle line remains close to a  Lorentzian  form with $\Gamma=-{\rm Im}\Sigma^{EEI}(k_F,0)-{\rm Im}\Sigma^{EPI}(0) +\Gamma_{imp}$.
A small constant impurity term, $\Gamma_{imp}$ has been added to the scattering rate.  At zero temperature and for $\omega=0$ we expect $\Gamma \cong \Gamma_{imp}$ and so the quasiparticle line width is reduced over its bare case by a factor of $1/Z^{tot}$ and the area under this part of the spectral density is reduced from a value of one by the same factor.
Note the narrowing of the main quasiparticle line in the interacting case as compared with the bare case, shown as the solid black curve.  The spectral weight lost in this region is transferred to EP and EE incoherent side bands which provide structures at higher energies ($\omega/\mu_0 > 0.2$) and make the interacting case decay less rapidly than does the bare case as $\omega/\mu_0$ increases.
    Note that the Lorentzian falls slightly below the interacting case even below $\omega/\mu_0 \equiv 0.2$.  This reflects the fact that the Coulomb side bands extend all the way down to $\omega=0$.
The missing spectral weight which is transferred to side bands is shown more clearly in Fig.~\ref{fig:2}(b) where we have an expanded view of the tail region showing the sidebands associated with the various interactions.
The solid black curve is the bare case as in Fig.~\ref{fig:2}(a) and is for comparison.  The curve containing both EEI and EPI, the red double-dashed-dotted curve, shows a Holstein type phonon assisted region with an onset at the phonon energy $\omega_E$.
To emphasize this particular feature a single $\delta$-function was used for the electron-phonon spectral density $\alpha^2F(\Omega)=\frac{\lambda \omega_E}{2}\delta\left(\Omega-\omega_E \right)$ with $\lambda$ set equal to $\approx 0.185$.  The blue dash dotted curve is for EPI alone and the green dashed curve is for EEI only.  These curves are quite distinct in that the EPI self energy shows a sharp rise at $\omega_E$ while the EEI case is much smoother in its behavior.  Nevertheless, it differs significantly from the black curve of the bare case.
In particular, the curves cross slightly above $\omega/\mu_0\cong 0.2$, in our normalized units, with the curve associated with the dressed case remaining substantially above the bare case.  It is precisely these deviations which tell us about interactions and which we will examine in the next section in much more detail using maximum entropy inversion techniques.  For now we point out only one feature.  The EEI and EPI together are not simply additive in the spectral function curves.

\begin{figure}
\vspace*{-0.5cm}%
\centerline{\includegraphics[width=3.5 in]{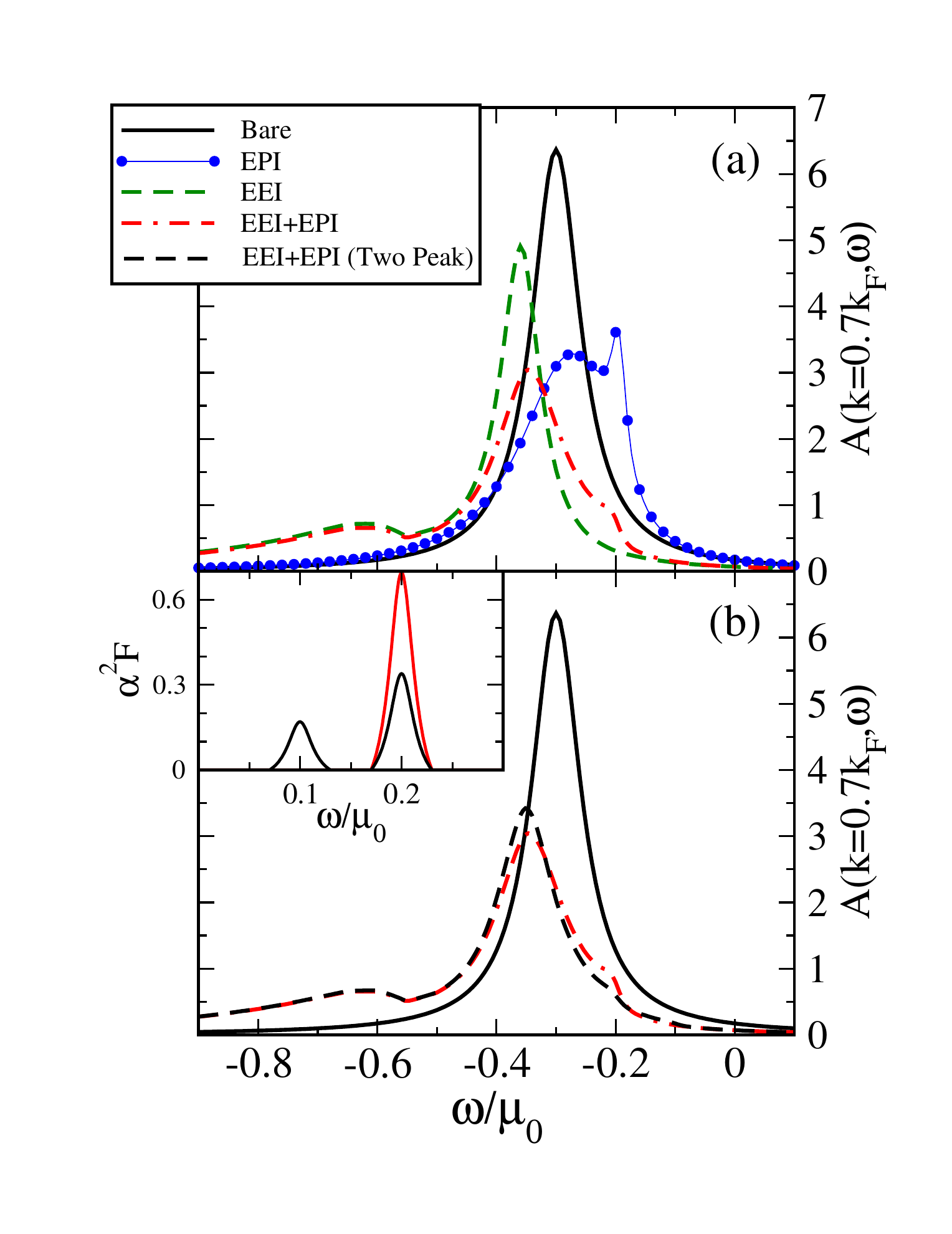}}
\vspace*{-1.0cm}%
  \caption{(Color online) (a) Electronic spectral density, $A(k=0.7k_F,\omega)$, at a momentum away from the Fermi momentum as a function of $\omega/\mu_0$ for the bare, EEI, EPI and combined cases.   (b) Electronic spectral density, $A(k=0.7k_F,\omega)$, at a momentum away from the Fermi momentum as a function of $\omega/\mu_0$.  The solid black curve is the bare case for comparison.  The red double-dashed-dotted line contains both EEI and EPI with a single peak electron-phonon spectral density $\alpha^2 F(\Omega)$ shown in the inset, while the green dashed line has a two peak $\alpha^2 F(\Omega)$ with the same mass enhancement $\lambda^{EPI}=0.2$ as the single peak spectrum.\label{fig:kp7inset} }
\end{figure}

The curves shown in Fig.~\ref{fig:2} are for $k=k_F$ and the structures due to EEI and EPI overlap in energy.
The situation for other values of momentum can be quite different.  This is illustrated in Fig.~\ref{fig:kp7inset} for $k=0.7k_F$.  In both the top and bottom frame the solid black curve is the bare case.  In the top frame the dashed green includes EEI only and the blue line with circles the EPI only.  These are to be compared with the red double-dashed-dotted curve which has both the EEI and EPI.  On comparing EPI alone with the bare case we see that the phonon structure appears (at $\omega=200$~meV) on the right hand side of the main quasiparticle line.  By contrast, the pure EEI shows a sideband on the left hand side of the remaining main line with only small distortion from a Lorentzian profile on the right.
Including the EPI does not alter the lower left part of the curves much but does introduce modifications to the right side.  Clearly in this case, the EEI and EPI have a separate energy window in which they dominate and this will be exploited in the next section to separate the two effects.  While we previously used a $\delta$-function in the EPI for illustration purposes, we have done further calculations with extended spectra as shown in  Fig.~\ref{fig:kp7inset}(b).  The double-dashed-dotted red is for a single truncated Lorentzian form and is illustrated in the inset.  The dashed green has instead two peaks, one centered at $0.1$~eV and the other centered at $0.2$~eV.
Differences associated with these two spectra can certainly be seen on the right hand side of the main quasiparticle line while the sideband associated with EEI instead remains the same.

It has become standard in the analysis of ARPES spectra to extract, from the energy distribution curves, information on the self energy as has been reviewed in Ref.~\onlinecite{carbotte:2011}.  A recent example applied to the high-T$_c$ cuprates is described in Zhang et al.\cite{zhang:2008} where detailed plots for the real part of the self energy are presented.  Of course the imaginary part also follows by Kramers-Kronig transforms of experimental data.

\begin{figure}
\vspace*{-0.5cm}%
\centerline{\includegraphics[width=3.5 in]{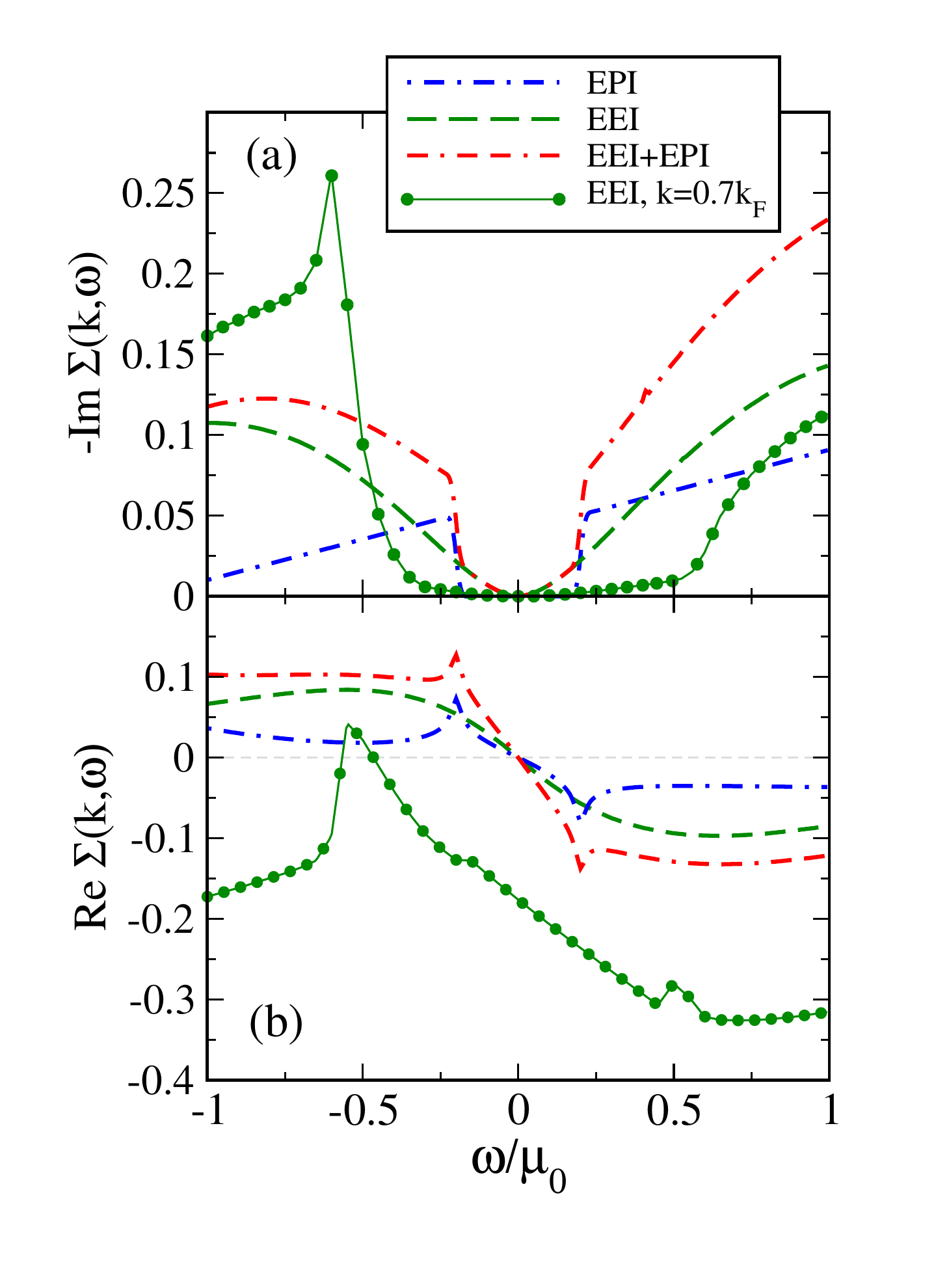}}
\vspace*{-1.0cm}%
  \caption{(Color online) (a) The negative of the imaginary part of the electronic self energy as a function of energy, $\omega/\mu_0$, at the Fermi momentum, $k_F$.  The green circles are for a case including only the EEI, but at a momentum away from the Fermi momentum ($k=0.7_F$) as in Fig.~\ref{fig:kp7inset}. (b) Same cases and notation as in (a) but now showing the real part of the electronic self energy.\label{fig:imagparts}}
\end{figure}

Our own results for real and imaginary parts of the EPI self energy, based on Eq.~(\ref{eqn:episigma}), are presented in Fig.~\ref{fig:imagparts}.  The dashed-dotted blue curves are for EPI only in a distributed phonon model with central frequency at $\omega_E=0.2$~eV.  In this case the imaginary part of the self energy is zero for $|\omega|<\omega_E$.  Crossing $\omega_E$ the curve rises to a finite value, after which it follows the dependence in energy of the bare density of states of graphene.  By contrast, for a conventional metal, the imaginary part of the self energy would simply be constant in this energy range, reflecting the constant DOS.
A second feature that needs to be mentioned is that the electron-phonon interaction is independent of the electron momentum in our model.  This is not the case for the electron-electron interactions based on Eq.~(\ref{eqn:eei}).  For $k=k_F$ the imaginary part of the self energy for positive $\omega$ grows gradually out of zero at $\omega=0$ [green dashed curve of Fig.~\ref{fig:imagparts}(a)] and reaches about half the value of the imaginary part of the EP self energy at $\omega=\omega_E$ after which it keeps growing with a change in slope from concave up to concave down.  On the other hand, for $k=0.7k_F$ the behavior is very different.  There is negligible EEI scattering for low energies which illustrates the idea that the EEI and EPI scatterings have their own separate windows away from the Fermi momentum while for $k=k_F$ they overlap and are not easily separated.  This will be more rigorously confirmed  in the next section when we employ maximum entropy inversion techniques to extract from such imaginary parts an effective electron-boson spectral function which describes the excitations responsible for the quasiparticle scattering.
The lower frame shows the corresponding real parts of the self energies.  These contain no independent information as they follow from the imaginary self energy through a Kramers-Kronig transformation.  We note one important feature of these curves.  They all go through zero at the Fermi energy for $k=k_F$.  Further, the slope out of $\omega=0$ gives the corresponding contribution to the velocity enhancement which is additive for the combined EEI+EPI cases.  This is also the case for the imaginary parts in Fig.~\ref{fig:imagparts}(a).

\begin{figure}
\vspace*{-0.5cm}%
\centerline{\includegraphics[width=3.4 in]{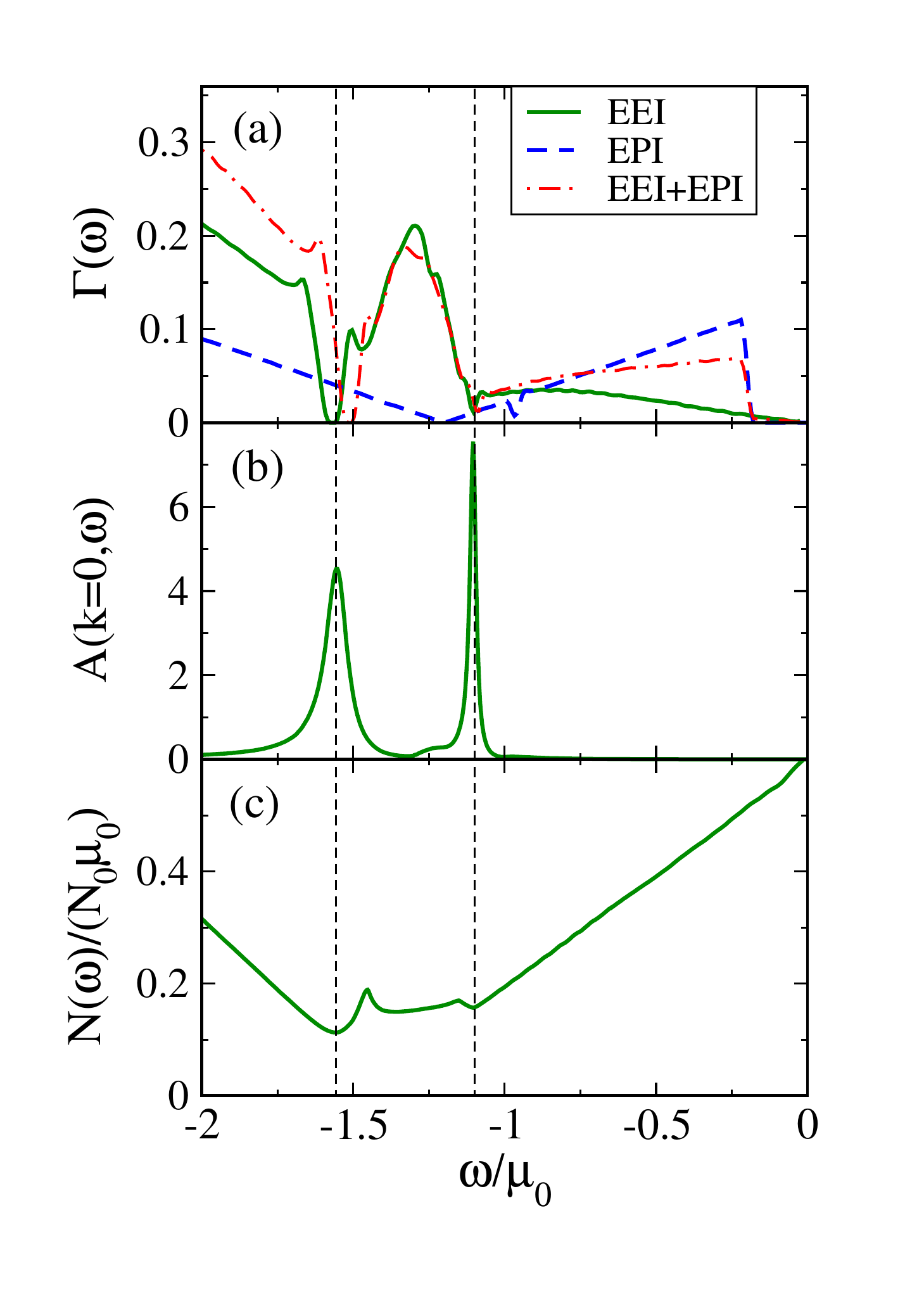}}
\vspace*{-0.8cm}%
  \caption{\label{fig:mdcwidths}(Color online) (a)MDC widths, $\Gamma(\omega)$, defined as the full width of the dominant peak in an MDC cut through the spectral function. (b) The spectral function, $A(k=0,\omega)$, through zero momentum including only the EEI. (c) Density of states for doped graphene including the EEI.  Note that the energy features of all three plots indicate the existence of two separate Dirac-like points along $k=0$.  }
\end{figure}

Before proceeding to a more detailed analysis of this data, we make a final point in Fig.~\ref{fig:mdcwidths} where we show our results for the MDC widths as a function of energy.  These are obtained from the numerical data of Fig.~\ref{fig:colourplot} by taking constant energy slices. Thus we define $\Gamma(\omega)$ to be the full width of the dominant spectral peak.  While the self energy associated with the EPI is strictly independent of k, the EEI is characterized as providing some momentum dependence to the renormalization as is seen clearly in Fig.~\ref{fig:colourplot}. Nevertheless we find that the MDCs have a well defined line width, $\Gamma(\omega)$. In Fig.~\ref{fig:mdcwidths}(a)  we compare the EPI (dashed blue), EEI (solid green) and EEI+EPI (dashed-dotted red) cases.  The EEI cases starts from zero at $\omega=0$ but rises at low frequencies to a finite value while the EPI contribution remains zero.  However, the EPI interaction dominates over the EEI contribution for a range of energies below $\omega_E=-0.2$~eV.  It is important to note that if a strongly distributed phonon spectrum had been used then the MDC line width would reflect this, and show finite values at frequencies between $-\omega_E$ and zero.  However, for $|\omega| > |\omega_E|$ there would be no difference between a delta function model with all the phonon spectral weight at $\omega_E$ and any arbitrarily distributed models with the same area under $\alpha^2F(\Omega)$ which cuts off at $\omega_E$.  Thus, unless this area is much smaller in reality than has been assumed in our model, the electron-phonon interactions are likely to contribute significantly to the low energy region of $\Gamma(\omega)$ although they cannot easily be disentangled from the EEI contribution in such a plot.

Turning to the sum of the two interactions (dashed-dotted red) we first note that this curve is not a simple addition of the two separate cases.  This would indeed be expected if only the imaginary part of the phonon self energy was considered in the spectral density of Eq.~\ref{eqn:akw}.  However, the real part also enters through the energy denominator and this leads to none additive features and to changes in $\Gamma(\omega)$ for both the low and high energy ranges.  The same lack of additivity is also seen at higher energies.  It is interesting to note that the reduction in $\Gamma(\omega)$ to near-zero values at $\omega \approx -1.1$ and $-1.56$ defines two well known points in the dressed dispersion curves; namely the two Dirac points at E$_0$ and E$_2$.\cite{bostwick:2010, leblanc:2011}
Although not a rigorous fit, our result for $\Gamma(\omega)$ in the G$_0$W-RPA calculation is consistent with the experimental ARPES data of Bostwick et al.\cite{bostwick:2007} and \emph{ab initio} simulations of the same spectra.\cite{park:2009:nl}
In Fig.~\ref{fig:mdcwidths}(b) and (c)  we show the spectral density for $k=0$, $A(k=0,\omega)$, and the electronic density of states (DOS), respectively, for the case of EEI.  The DOS is obtained by summing the spectral density over the Brillouin zone and results in a quantity which also shows minima at E$_0$ and E$_2$ that align precisely with the peaks in $A(k=0,\omega)$ and corresponding minima in $\Gamma(\omega)$.\cite{leblanc:2011,principi:2011}

\section{Maximum Entropy Inversions}\label{sec:inversion}

The self energy for an interacting Dirac fermion which we denote by $\Sigma(k,\omega)$ is given by Eq.~(\ref{eqn:eei}) which provides the electron-electron contribution.  We include an additional electron-phonon contribution given by Eq.~(\ref{eqn:episigma}).  For the pupose of performing a maximum entropy inversion, we denote the values of the total self energy by $D_i$ at a discrete set of frequencies, $\omega_i$, with $i=1,\cdot\cdot\cdot,N$ where $N$ is a large integer.  These discrete sets of frequencies and self energies serve as input data for the inversion\cite{schachinger:2006, schachinger:2008, wu:2010, bok:2010, carbotte:2011} of an appropriately chosen convolution integral assumed to represent the relationship between the total self energy, $\Sigma_{tot}(\omega)$, and an effective electron-boson spectral density, $\alpha^2F(\omega)$, which are related through
\begin{equation}
\Sigma_{tot}(\omega)=\int_0^\infty K(\omega,\Omega)\alpha^2F(\Omega)d\Omega.\label{eqn:inversiona}
\end{equation}
For the electron-phonon interaction alone, the relationship of Eq.~(\ref{eqn:inversiona}) is exact and the kernel, $K(\omega,\Omega)$, at zero temperature reduces to
\begin{widetext}
\begin{equation}
K(\omega,\Omega)=\int_{-\infty}^\infty d\omega^\prime \frac{|  \omega^\prime +\mu_0 |}{\mu_0}\left\{ \frac{\Theta(-\omega^\prime)}{\omega-\Omega-\omega^\prime+i0^+} +\frac{\Theta(\omega^\prime)}{\omega+\Omega-\omega^\prime+i0^+} \right\}.\label{eqn:inversionb}
\end{equation}
\end{widetext}
In practice, an appropriately chosen cutoff energy $W_c$ is to be applied to the integrals in Eq.~(\ref{eqn:inversionb}) to assure convergence.  For the electron-electron case, application of Eq.~(\ref{eqn:inversiona}) and (\ref{eqn:inversionb})  to represent the self energy implies the additional assumption that the results of Eqs.~(\ref{eqn:res}) and (\ref{eqn:line}) can be modelled by some effective electron-boson spectral density describing new boson exchange processes originating in the electron-hole excitations, plasmons or plasmarons.  The self energy components of Eq.~(\ref{eqn:eei}) do not rigorously map onto the mathematical form implied in Eqs.~(\ref{eqn:inversiona}) and (\ref{eqn:inversionb}) but, as we will see here, such an analysis is still very useful and valuable.  It allows us to identify the effect of the electron-electron interaction as a background to the electron-phonon interaction spectral density.

Given the input data, $D_i$, on a frequency grid, $\omega_i$, we want to find the corresponding $\alpha^2F(\Omega_i)\equiv \alpha^2 F_i$ of Eq.~(\ref{eqn:inversiona}). In discrete form
\begin{equation}
D_i=\sum_{j=1}^N K(\omega_i,\Omega_j)\alpha^2F(\Omega_j) \Delta\Omega_j \label{eqn:inversionc}
\end{equation}
where $\Delta\Omega_j$ is the grid size in the $\Omega$ integration.
This deconvolution problem is ill conditioned and here will be accomplished with the use of a maximum entropy method.\cite{carbotte:2011,schachinger:2006,schachinger:2008,wu:2010,bok:2010}
We define
\begin{equation}
\chi^2=\sum_{i=1}^N \frac{[D_i-\Sigma_{tot}(\omega_i)]^2}{\sigma_i^2}
\end{equation}
where $\Sigma_{tot}(\omega_i)$ is given by Eq.~(\ref{eqn:inversiona}) and is a functional of the effective electron-boson spectral density, $\alpha^2F(\Omega_j)$, that we wish to determine.  Here $\sigma_i$ is the error assigned to the data, $D_i$.  A least squares method would minimize $\chi^2$ against $\alpha^2 F(\Omega)$.  Of course, one needs to incorporate into the process some physical constraints such as the positive definite nature of $\alpha^2F(\omega)$ as it is representative of a boson exchange interaction.  Here we use a maximum entropy method in which the functional,
\begin{equation}\label{eqn:inversione}
L=\frac{\chi^2}{2}-aS,
\end{equation}
is minimized with $S$, the Shannon-Jones entropy, given as\cite{schachinger:2008}
\begin{equation}\label{eqn:inversionf}
S=\int_0^\infty \left[ \alpha^2F(\omega)-m(\omega)-\alpha^2F(\omega)\ln\left\{    \frac{\alpha^2F(\omega)}{m(\omega)}     \right\}        \right]d\omega.
\end{equation}
The parameter $m(\omega)$ is taken here to be some small constant value $m_0$ which corresponds to an initial assumption that we have no a priori knowledge of the functional form of the electron-boson spectral density.  Should we have such information, it could be used to initialize $m(\omega)$ to a form representative of the additional information on $\alpha^2F(\omega)$.  The parameter $a$ in Eq.~(\ref{eqn:inversione}) is a determinative parameter which controls how close the fitting follows the data.  We iterate on this parameter until the $\Sigma_{tot}(\omega_i)$ of Eq.~(\ref{eqn:inversiona}) are within $\sigma_i$ of the data points $D_i$.  More details can be found in Refs.~\onlinecite{schachinger:2006} and \onlinecite{carbotte:2011} where many cases of inversion of ARPES, optical conductivity and Raman data are reviewed.  In all of these previous cases, however, the inversion proceeds on the assumption that the electronic density of states around the Fermi surface does not vary significantly on the energy scale associated with the boson exchange processes involved.  The density of state factor $|\omega^\prime+\mu_0|/\mu_0$ in Eq.~(\ref{eqn:inversionb}) is replaced by its value at the Fermi energy where $\omega^\prime=0$.  Here we invert with the proper bare DOS factor that is linear in energy with a Dirac point at $-\mu_0$.  This will be referred to as the bare density of states (BDOS) case while the other will be referred to as the flat density of states case (FDOS) which we will present only for comparison.

\begin{figure}
\vspace*{-0.5cm}%
\centerline{\includegraphics[width=3.0 in, clip, trim= 13mm 26mm 25mm 25mm]{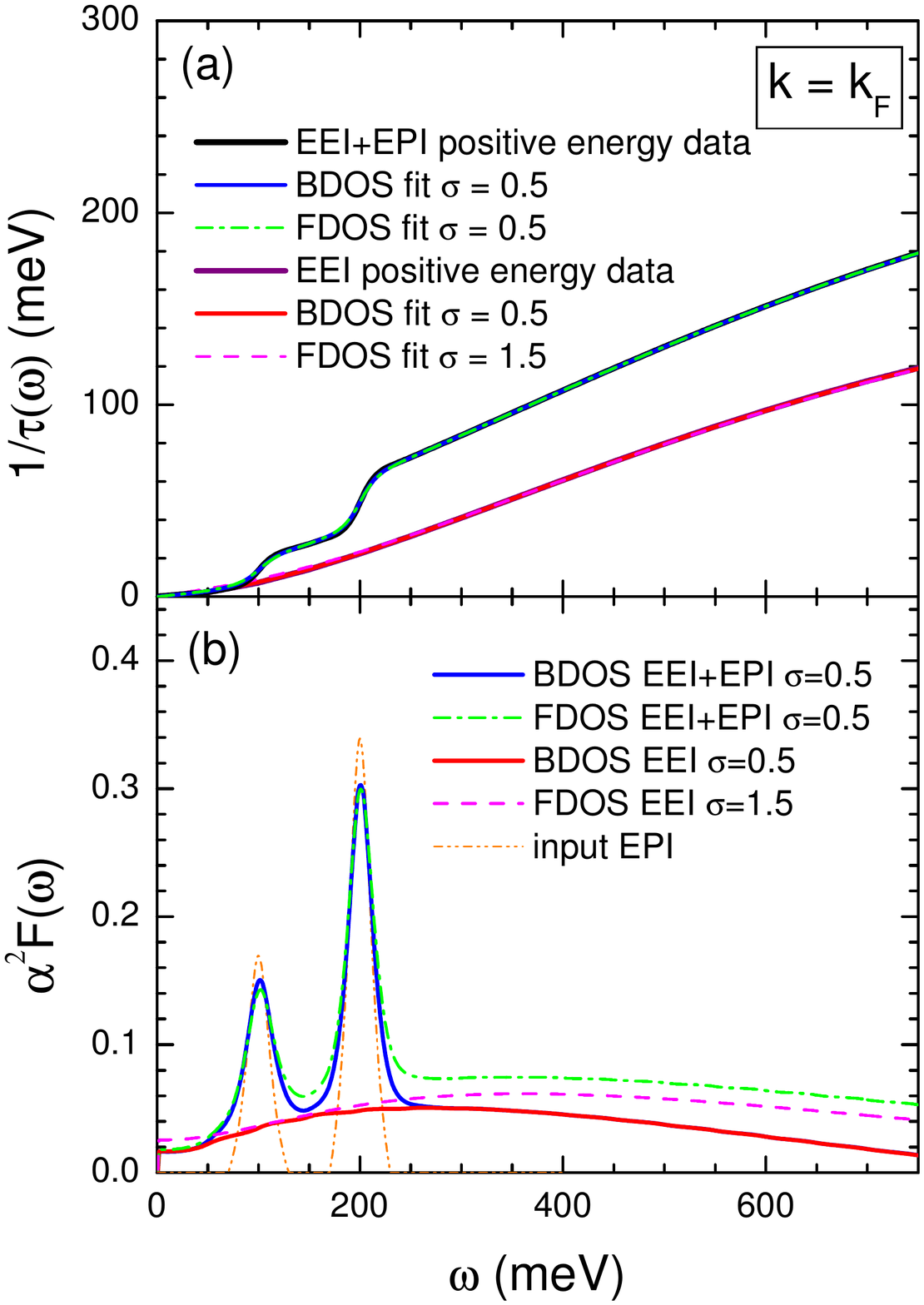}}
\vspace*{-0.0cm}%
  \caption{\label{fig:fig1_jh}(Color online) (a)~Quasiparticle scattering rate, $1/2\tau^{qp}(\omega)\equiv -{\rm Im}\Sigma(k,\omega)$, at the Fermi momentum, $k=k_F$, for the EEI and EEI+EPI self energies at positive $\omega$ along with resulting inversion fits using bare and flat density of states inversions (BDOS and FDOS respectively). (b)~Recovered effective electron boson spectral density, $\alpha^2F(\omega)$, from the $1/\tau(\omega)$ of (a).  The solid red and blue lines are the final fits to input data obtained using Eq.~(\ref{eqn:inversiona}) and self energy of Eq.~(\ref{eqn:inversionb}) while the dashed purple and dashed-dotted green were obtained from Eq.~(\ref{eqn:inversiona}) but where $|\omega^\prime+\mu_0|/\mu_0=1$.  The dashed-double-dotted orange curve is the input two-peak EPI $\alpha^2F(\omega)$.}
\end{figure}

In Fig.~\ref{fig:fig1_jh} we show results for the momentum at $k=k_F$.  Here  the positive energy range for the self energy data, solid black curve, includes the EEI along with an EPI part where $\lambda^{EPI}= 0.185$ and is comprised of two truncated Lorentzians as shown in the inset of Fig.~\ref{fig:kp7inset}.  The lower peak is centered at $100$~meV and the upper at $200$~meV.  This information is not used in any way in the inversion process based on Eqs.~(\ref{eqn:inversiona}) and (\ref{eqn:inversionb}) which proceeds on the assumption that nothing is known about the electron-boson functions that we seek to obtain from the electron spectral density data at $k=k_F$.  Here we compare the quality of the fit obtained using Eq.~(\ref{eqn:inversionb}) with the BDOS and FDOS cases.  We see that while $\sigma$ is larger in the second case, the overall fit remains good.  Note that the two steps associated with the phonon spectrum used are clearly seen and are well fit.  Also shown are results of a maximum entropy inversion for the case when only the EEI is included (solid purple curve).  In this instance the data for the quasiparticle lifetime $1/\tau(\omega)$ are comparatively smoother.  These are based on Eq.~(\ref{eqn:eei}) and consequently there is no rigorous reason that a boson exchange model, which is embodied in Eq.~(\ref{eqn:inversionb}), should properly describe the data.  However, we see an excellent inversion fit and recover a smooth bosonic spectrum shown in Fig.~\ref{fig:fig1_jh}(b) for the BDOS case (solid red).  This clearly demonstrates that a boson exchange model with a small, smooth effective bosonic spectral density, spread over a large energy scale of a few hundred meV, can reproduce the effect of the EEI on the quasiparticle spectrum.  When a phonon piece is added, as in the solid blue curve of Fig.~\ref{fig:fig1_jh}(b), we see prominent features of the two truncated Lorentzian peaks (similar to the input EPI) superimposed on the EEI background.  Both parts are easily recognizable as they have quite distinct behavior but in principle, they are hard to separate out precisely because the EEI and EPI spectra overlap in energy.
The dashed (purple) and dashed-dotted (green) curves are similar, but are obtained through the inversion for the FDOS case.  These are shown only for comparison and serve to show that to get a quantitative electron-boson exchange spectral density, $\alpha^2F(\omega)$, it is essential to include the correct linear in energy dependence of the bare graphene DOS.  While we still recover a reasonable qualitative representation of the actual spectrum of excitations involved in the quasiparticle scattering, the tails above $\approx 200$~meV are strongly overestimated when we use a flat density of states.

\begin{figure}
\vspace*{-0.0cm}%
\centerline{\includegraphics[width=3.5 in, trim= 28mm 25mm 30mm 25mm]{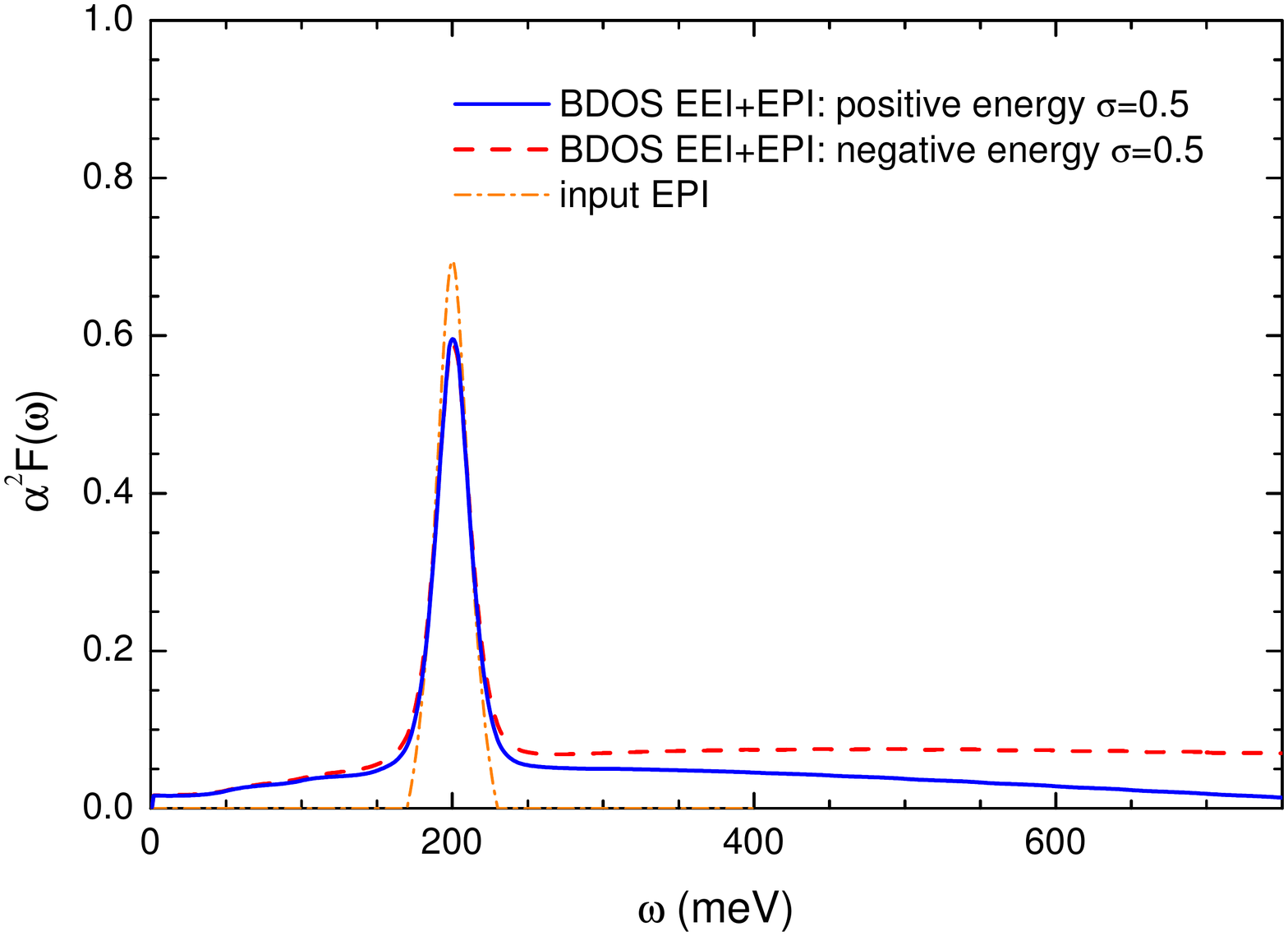}}
\vspace*{0.3cm}%
  \caption{\label{fig:fig2_jh}(Color online) Recovered effective boson spectral density $\alpha^2F(\omega)$ for the case of EEI~+~EPI using the positive (blue) and negative (red) energy parts of $-{\rm Im}\Sigma(k,\omega)$ for the inversion.  The dash dotted red curve is the input EPI part.  However, the EEI background is not symmetric about $\omega=0$.}
\end{figure}

From an examination of the data in Fig.~\ref{fig:imagparts} for the quasiparticle self energies it is evident that positive and negative energy results are not symmetric about the Fermi surface.  This is expected since the BDOS itself does not posses this symmetry due to the finite value of $\mu_0$.  Thus, the inversion of such data will produce different effective electron-boson exchange spectra.  This is illustrated in the results for the recovered spectra in Fig.~\ref{fig:fig2_jh} at $k=k_F$.  The solid blue curve gives our results when the positive $\omega$ data is used and the dashed-red is for the negative energy data.  The curves agree in the region of the phonon peak, here taken as a single truncated Lorentzian centered at 200~meV and shown as the dashed-dotted orange curve.  The EEI background is quite different for positive and negative energies.  It is important to note in these comparisons that $\alpha^2F(\omega)$ is a dimensionless quantity such that our results provide a true measure of the magnitude of the electron-boson exchange with no adjustable scale of any kind.

\begin{figure}
\vspace*{-0.0cm}%
\centerline{\includegraphics[width=3.5 in, trim= 28mm 25mm 30mm 25mm]{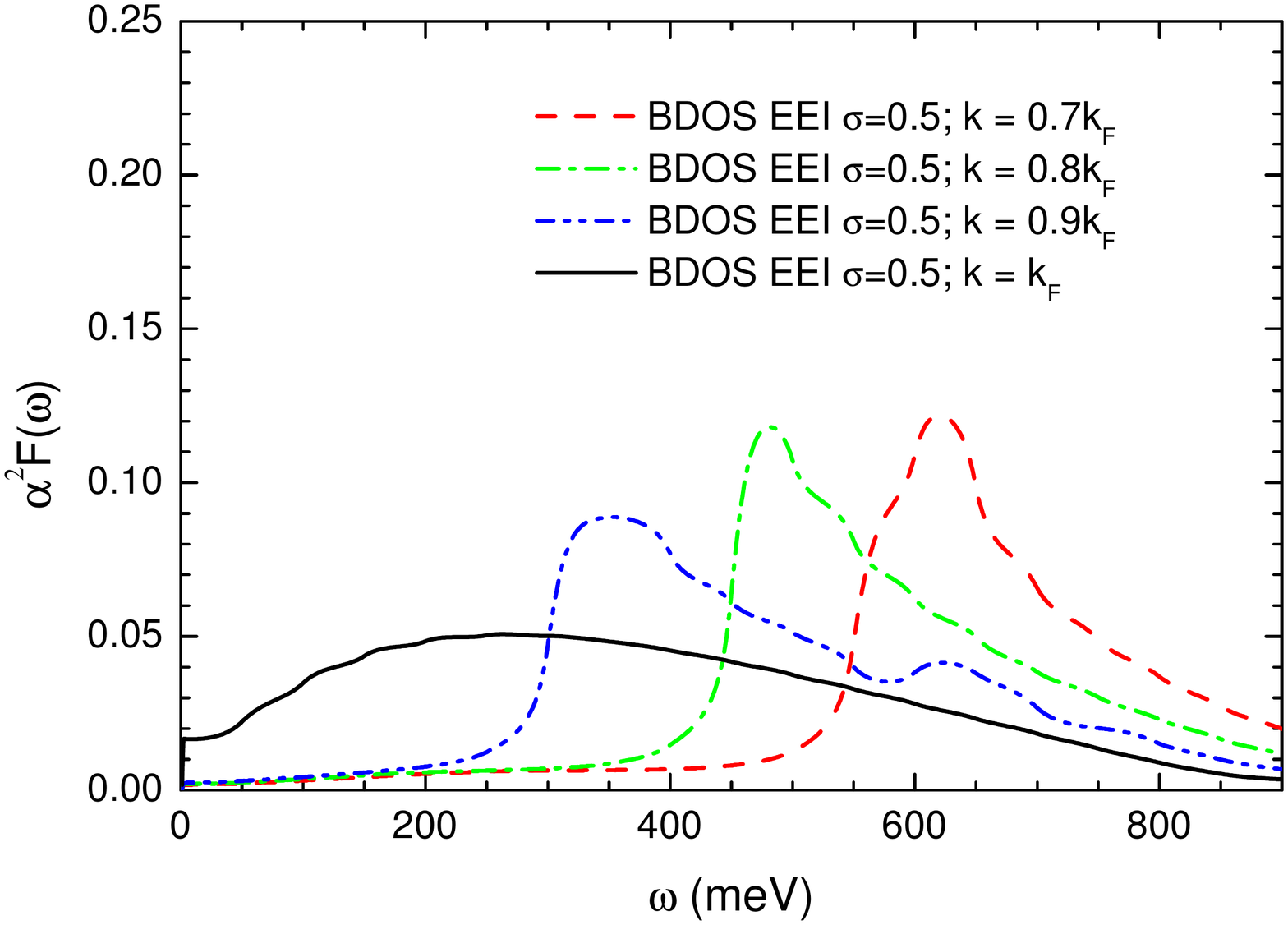}}
\vspace*{-0.0cm}%
  \caption{\label{fig:newfig_jh}(Color online) Clearing of electron-phonon energy window of all electron-electron contributions by decreasing momentum below $k_F$.  Data is shown for $k=0.7$, 0.8, 0.9, and 1.0 times the Fermi momentum value, $k_F$.  A region of nearly zero value forms in the effective spectral density at small $\omega$ for $k \neq k_F$.}
\end{figure}

It is clear from examination of the self energy due to the EEI for $k=k_F$ and $k=0.7k_F$ (Fig.~\ref{fig:imagparts}) that these renormalizations depend strongly on the value of the momentum.  More important is the nearly zero value of the imaginary part of the self energy and the observed asymmetry  illustrated by the  $k=0.7k_F$ case where the self energy is close to zero up to $0.5$ on the positive frequency side and approximately $0.3$ on the negative side. This clearing of the low energy region of electron-electron renormalization provides a window of opportunity to display on their own the electron-phonon renormalizations.  In Fig.~\ref{fig:newfig_jh} we show the evolution of the recovered effective $\alpha^2F(\omega)$ for the case of EEI alone at several values of momentum ranging between $k=k_F$ and $k=0.7k_F$.  It is clear that as we move away from the Fermi momentum $\alpha^2F(\omega)$ rapidly develops a gap at low energy which increases with decreasing value of $k$.  This is the region in which an addition of an electron-phonon part to the spectral density will fall.  Thus it will show up without a background.

\begin{figure}
\vspace*{-0.0cm}%
\centerline{\includegraphics[width=3.0 in, trim= 10mm 26mm 25mm 25mm]{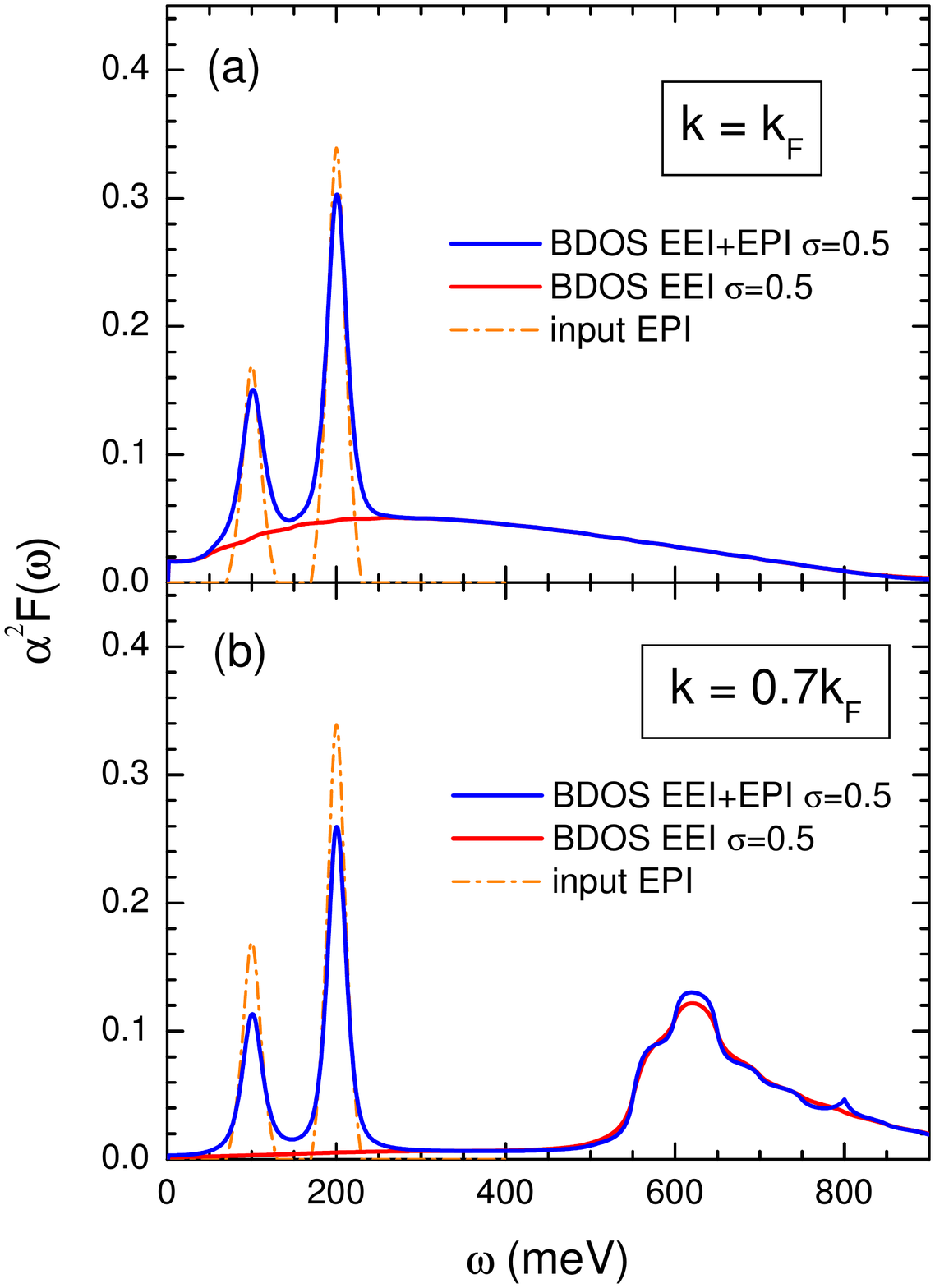}}
\vspace*{-0.0cm}%
  \caption{\label{fig:fig3_jh}(Color online) Recovered effective electron boson spectral density $\alpha^2F(\omega)$ from the quasiparticle self energy at: (a) $k=k_F$ and (b) $k=0.7k_F$.  The blue curve includes both the EEI background and an EPI part shown as the two truncated Lorentzians as the dash-dotted orange curve.  The red curve is the results when only EEI are included.}
\end{figure}

In Fig.~\ref{fig:fig3_jh} we show results for the combined EEI and EPI  when we move away from the Fermi momentum.  Here, the recovered spectrum for $k=k_F$ is shown in Fig.~\ref{fig:fig3_jh}(a) while the one for $k=0.7k_F$ is shown on the same scale for easy comparison in Fig.~\ref{fig:fig3_jh}(b).  Several features need to be emphasized.  First the electron-phonon spectra is now revealed in its own energy window.  In the energy range of the phonons in our model ($\omega<200$meV) the electron-electron contribution to $\alpha^2F(\omega)$ is negligible.  This is different from the case of $k=k_F$, where they provide a significant background which would need to be subtracted out in some way for the phonon spectrum to be revealed on its own.  Another important feature of the spectrum is that electron-electron interactions do not contribute significantly until $\omega>500$meV for the case chosen where $\mu_0=400$meV.  This is completely different from the $k=k_F$ case where the EEI background is nonzero all the way down to $\omega=0$ and shows a maximum around $300$meV.  It is never large as compared with the $k=0.7k_F$ case.
In that case, the large electron-electron peak extending from $550-800$~meV is associated with the plasmaron structures which develop between $\omega=1.2\mu_0 \rightarrow 2 \mu_0$  described in the previous section and is a good indication of the strength of the EEI.    To see sharp features of the plasmaron structure in the recovered $\alpha^2F(\omega)$ one needs to sample momenta below $k=k_F$.  This result of the inversion process is consistent with spectral function presented in Fig.~\ref{fig:colourplot}.

\begin{figure}
\vspace*{-0.0cm}%
\centerline{\includegraphics[width=3.0 in, trim= 10mm 26mm 25mm 25mm]{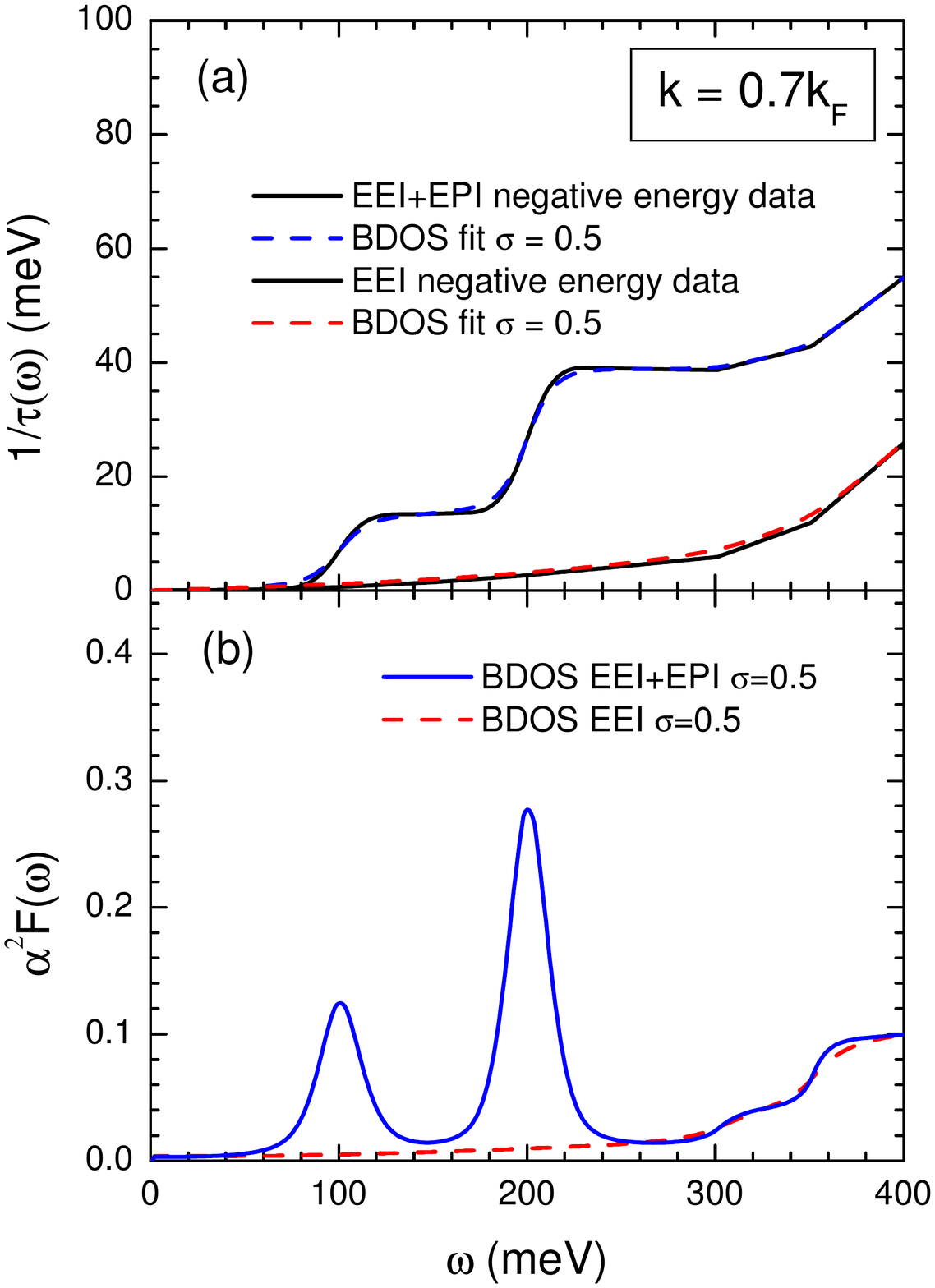}}
\vspace*{-0.0cm}%
  \caption{\label{fig:fig10_jh}(Color online) (a) Quasiparticle scattering rate away from the Fermi momentum at $k=0.7k_F$, for EEI and EEI+EPI self energies at negative $\omega$ along with their bare density of states (BDOS) inversion fits. (b) Recovered effective electron-boson spectral density, $\alpha^2F(\omega)$, from the $1/\tau(\omega)$ of (a).  Unlike the $k=k_F$ case of Fig.~\ref{fig:fig1_jh}, the low frequency window contains negligible EEI contamination resulting in bosonic features due only to the EPI.}
\end{figure}

Fig.~\ref{fig:fig3_jh} was obtained through the inversion of positive frequency data for the self energy.  This was done because in this case the electron-electron structure is displaced from zero to energies well above the phonon cutoff of 200~meV.  Of course, ARPES only provides information about the negative frequency range of the self energy.  There is no particle-hole symmetry and Coulomb interaction effects are seen to start in the self energy for $k=0.7k_F$ and at negative energies at around $\approx -300$~meV as seen in Fig.~\ref{fig:imagparts}.  Nevertheless, even in this case the electron-phonon spectrum is largely independent of any contamination by electron-electron interactions.  The imaginary part of the self energy is shown in Fig.~\ref{fig:fig10_jh}(a) (here in the form of $1/\tau(\omega)$) for $k=0.7k_F$.  Examples are shown for EEI only and for EEI + EPI with the EPI part based on a two truncated Lorentzian model for the electron-phonon $\alpha^2F(\omega)$ as in Fig.~\ref{fig:fig1_jh}.  The dashed curves are the maximum entropy inversion fits to the data.  In the lower frame we compare the recovered spectral densities for the two cases.  We see  that the electron-phonon contribution is distinct from any significant EE contribution which only has strong features at higher energies above approximately 300~meV which is well beyond the phonon cutoff.  It is the stark contrast between the $k=k_F$ case of Fig.~\ref{fig:fig1_jh} and the $k=0.7k_F$ case shown in Fig.~\ref{fig:fig3_jh} that should impress upon the reader the importance of the finite $k$ analysis for elucidating the precise $\alpha^2F(\omega)$ of the EPI in graphene that one can obtain through inversion of ARPES data.

\section{Conclusions}\label{sec:conclusions}

We have calculated the effects of the electron-electron correlations, within the G$_0$W-RPA approximation, and also of the electron-phonon interaction on the spectral function of the massless Dirac charge carriers of graphene.  The aim is to examine the combined signatures of these two renormalization processes that are encoded in the self energies and related quantities, with a view at finding a way to separate the two contributions and as a result isolate information on each individually.  For momentum equal to the Fermi momentum, the sidebands on either side of the main quasiparticle peak reflect contributions from both EEI and EPI.  Their individual contributions are hard to separate as they overlap in energy.  Nevertheless, they differ qualitatively and in that sense can be distinguished somewhat.  The EEI provides a smooth background extending over an energy scale of the order of the chemical potential $\mu_0$ which in this work is taken to be variable, but of an experimentally relevant range from $0.1 \to 1$~eV.  By contrast, the electron-phonon interaction provides a peak structure reflecting the various phonon branches and cuts off at the Debye energy of order $\omega_D=200$~meV in graphene.  A clearer image of the two separate interactions can be obtained by going to lower momentum values.  For the specific case of $k=0.7k_F$ the wings on the electron spectral density on the right hand side of the main quasiparticle peak are due entirely to the interaction with phonons.  The EE contribution has instead moved entirely into the left sideband which is itself hardly affected by the phonon interaction.  Following what is done in dealing with experimental angular resolved photoemission data in many other materials we use a maximum entropy inversion technique to relate our numerical data on the quasiparticle self energy at momentum, $k$, as a function of energy (which in experiment is derived from the ARPES data on the spectral density) to an effective electron boson spectral density denoted by $\alpha^2 F(\omega)$.  This procedure shows clearly how for $k=k_F$, the phonon self energy is embedded into an electron-electron background, while for $k=0.7k_F$ it shows up separately in its own energy window with negligible electron-electron background.  There the background is only significant for higher energies; well above the phonon cut off.  Beyond this, a peak is seen in the recovered effective density which is related to plasmaron structures which have their origin in the EEI.  Plasmarons are also expected to exist in graphene's interacting density of electronic states, $N(\omega)$, as discussed recently in Refs.~\onlinecite{leblanc:2011, principi:2011}.  Here we find that related structure exists in the widths, $\Gamma(\omega)$, of the momentum distribution curves defined in ARPES.  These structures are shifted by the introduction of the electron-phonon interaction but fall in a different energy range from the more direct image of the phonons at low energy below $\omega_D$.


\begin{acknowledgments}
  J.H. acknowledges financial support from the National Research Foundation of Korea (NRFK Grant No. 20100008552).  J.C. acknowledges support from the Natural Sciences and
Engineering Research Council of Canada  and the Canadian Institute
for Advanced Research.   We thank Prof. E. Schachinger for providing a maximum entropy routine and for his assistance in modifying it to deal with the cases considered here.
\end{acknowledgments}


\bibliographystyle{apsrev4-1}
\bibliography{bib}

\begin{thebibliography}{58}%
\makeatletter
\providecommand \@ifxundefined [1]{%
 \@ifx{#1\undefined}
}%
\providecommand \@ifnum [1]{%
 \ifnum #1\expandafter \@firstoftwo
 \else \expandafter \@secondoftwo
 \fi
}%
\providecommand \@ifx [1]{%
 \ifx #1\expandafter \@firstoftwo
 \else \expandafter \@secondoftwo
 \fi
}%
\providecommand \natexlab [1]{#1}%
\providecommand \enquote  [1]{``#1''}%
\providecommand \bibnamefont  [1]{#1}%
\providecommand \bibfnamefont [1]{#1}%
\providecommand \citenamefont [1]{#1}%
\providecommand \href@noop [0]{\@secondoftwo}%
\providecommand \href [0]{\begingroup \@sanitize@url \@href}%
\providecommand \@href[1]{\@@startlink{#1}\@@href}%
\providecommand \@@href[1]{\endgroup#1\@@endlink}%
\providecommand \@sanitize@url [0]{\catcode `\\12\catcode `\$12\catcode
  `\&12\catcode `\#12\catcode `\^12\catcode `\_12\catcode `\%12\relax}%
\providecommand \@@startlink[1]{}%
\providecommand \@@endlink[0]{}%
\providecommand \url  [0]{\begingroup\@sanitize@url \@url }%
\providecommand \@url [1]{\endgroup\@href {#1}{\urlprefix }}%
\providecommand \urlprefix  [0]{URL }%
\providecommand \Eprint [0]{\href }%
\providecommand \doibase [0]{http://dx.doi.org/}%
\providecommand \selectlanguage [0]{\@gobble}%
\providecommand \bibinfo  [0]{\@secondoftwo}%
\providecommand \bibfield  [0]{\@secondoftwo}%
\providecommand \translation [1]{[#1]}%
\providecommand \BibitemOpen [0]{}%
\providecommand \bibitemStop [0]{}%
\providecommand \bibitemNoStop [0]{.\EOS\space}%
\providecommand \EOS [0]{\spacefactor3000\relax}%
\providecommand \BibitemShut  [1]{\csname bibitem#1\endcsname}%
\let\auto@bib@innerbib\@empty
\bibitem [{\citenamefont {Geim}\ and\ \citenamefont
  {Novoselov}(2007)}]{geim:2007}%
  \BibitemOpen
  \bibfield  {author} {\bibinfo {author} {\bibfnamefont {A.~K.}\ \bibnamefont
  {Geim}}\ and\ \bibinfo {author} {\bibfnamefont {K.~S.}\ \bibnamefont
  {Novoselov}},\ }\href@noop {} {\bibfield  {journal} {\bibinfo  {journal}
  {Nat. Mat.}\ }\textbf {\bibinfo {volume} {6}},\ \bibinfo {pages} {183}
  (\bibinfo {year} {2007})}\BibitemShut {NoStop}%
\bibitem [{\citenamefont {{Castro Neto}}\ \emph {et~al.}(2009)\citenamefont
  {{Castro Neto}}, \citenamefont {Guinea}, \citenamefont {Peres}, \citenamefont
  {Novoselov},\ and\ \citenamefont {Geim}}]{neto:2009}%
  \BibitemOpen
  \bibfield  {author} {\bibinfo {author} {\bibfnamefont {A.~H.}\ \bibnamefont
  {{Castro Neto}}}, \bibinfo {author} {\bibfnamefont {F.}~\bibnamefont
  {Guinea}}, \bibinfo {author} {\bibfnamefont {N.~M.~R.}\ \bibnamefont
  {Peres}}, \bibinfo {author} {\bibfnamefont {K.~S.}\ \bibnamefont
  {Novoselov}}, \ and\ \bibinfo {author} {\bibfnamefont {A.~K.}\ \bibnamefont
  {Geim}},\ }\href@noop {} {\bibfield  {journal} {\bibinfo  {journal} {Rev.
  Mod. Phys.}\ }\textbf {\bibinfo {volume} {81}},\ \bibinfo {pages} {109}
  (\bibinfo {year} {2009})}\BibitemShut {NoStop}%
\bibitem [{\citenamefont {Gusynin}\ \emph {et~al.}(2007)\citenamefont
  {Gusynin}, \citenamefont {Sharapov},\ and\ \citenamefont
  {Carbotte}}]{gusynin:2007:ijmp}%
  \BibitemOpen
  \bibfield  {author} {\bibinfo {author} {\bibfnamefont {V.~P.}\ \bibnamefont
  {Gusynin}}, \bibinfo {author} {\bibfnamefont {S.~G.}\ \bibnamefont
  {Sharapov}}, \ and\ \bibinfo {author} {\bibfnamefont {J.~P.}\ \bibnamefont
  {Carbotte}},\ }\href@noop {} {\bibfield  {journal} {\bibinfo  {journal}
  {International Journal of Modern Physics B}\ }\textbf {\bibinfo {volume}
  {21}},\ \bibinfo {pages} {4611} (\bibinfo {year} {2007})}\BibitemShut
  {NoStop}%
\bibitem [{\citenamefont {Abergel}\ \emph {et~al.}(2010)\citenamefont
  {Abergel}, \citenamefont {Apalkov}, \citenamefont {Berashevich},
  \citenamefont {Ziegler},\ and\ \citenamefont {Chakraborty}}]{abergel:2010}%
  \BibitemOpen
  \bibfield  {author} {\bibinfo {author} {\bibfnamefont {D.~S.~L.}\
  \bibnamefont {Abergel}}, \bibinfo {author} {\bibfnamefont {V.}~\bibnamefont
  {Apalkov}}, \bibinfo {author} {\bibfnamefont {J.}~\bibnamefont
  {Berashevich}}, \bibinfo {author} {\bibfnamefont {K.}~\bibnamefont
  {Ziegler}}, \ and\ \bibinfo {author} {\bibfnamefont {T.}~\bibnamefont
  {Chakraborty}},\ }\href@noop {} {\bibfield  {journal} {\bibinfo  {journal}
  {Adv. Phys.}\ }\textbf {\bibinfo {volume} {59}},\ \bibinfo {pages} {261}
  (\bibinfo {year} {2010})}\BibitemShut {NoStop}%
\bibitem [{\citenamefont {Orlita}\ and\ \citenamefont
  {Potemski}(2010)}]{orlita:2010}%
  \BibitemOpen
  \bibfield  {author} {\bibinfo {author} {\bibfnamefont {M.}~\bibnamefont
  {Orlita}}\ and\ \bibinfo {author} {\bibfnamefont {M.}~\bibnamefont
  {Potemski}},\ }\href@noop {} {\bibfield  {journal} {\bibinfo  {journal}
  {Semiconductor Science Technology}\ }\textbf {\bibinfo {volume} {25}},\
  \bibinfo {pages} {063001} (\bibinfo {year} {2010})}\BibitemShut {NoStop}%
\bibitem [{\citenamefont {Bostwick}\ \emph {et~al.}(2007)\citenamefont
  {Bostwick}, \citenamefont {Ohta}, \citenamefont {Seyller}, \citenamefont
  {Horn},\ and\ \citenamefont {Rotenberg}}]{bostwick:2007}%
  \BibitemOpen
  \bibfield  {author} {\bibinfo {author} {\bibfnamefont {A.}~\bibnamefont
  {Bostwick}}, \bibinfo {author} {\bibfnamefont {T.}~\bibnamefont {Ohta}},
  \bibinfo {author} {\bibfnamefont {T.}~\bibnamefont {Seyller}}, \bibinfo
  {author} {\bibfnamefont {K.}~\bibnamefont {Horn}}, \ and\ \bibinfo {author}
  {\bibfnamefont {E.}~\bibnamefont {Rotenberg}},\ }\href@noop {} {\bibfield
  {journal} {\bibinfo  {journal} {Nat. Phys.}\ }\textbf {\bibinfo {volume}
  {3}},\ \bibinfo {pages} {36} (\bibinfo {year} {2007})}\BibitemShut {NoStop}%
\bibitem [{\citenamefont {Bianchi}\ \emph {et~al.}(2010)\citenamefont
  {Bianchi}, \citenamefont {Rienks}, \citenamefont {Lizzit}, \citenamefont
  {Baraldi}, \citenamefont {Balog}, \citenamefont {Hornek{\ae}r},\ and\
  \citenamefont {Hofmann}}]{bianchi:2010}%
  \BibitemOpen
  \bibfield  {author} {\bibinfo {author} {\bibfnamefont {M.}~\bibnamefont
  {Bianchi}}, \bibinfo {author} {\bibfnamefont {E.~D.~L.}\ \bibnamefont
  {Rienks}}, \bibinfo {author} {\bibfnamefont {S.}~\bibnamefont {Lizzit}},
  \bibinfo {author} {\bibfnamefont {A.}~\bibnamefont {Baraldi}}, \bibinfo
  {author} {\bibfnamefont {R.}~\bibnamefont {Balog}}, \bibinfo {author}
  {\bibfnamefont {L.}~\bibnamefont {Hornek{\ae}r}}, \ and\ \bibinfo {author}
  {\bibfnamefont {P.}~\bibnamefont {Hofmann}},\ }\href@noop {} {\bibfield
  {journal} {\bibinfo  {journal} {Phys. Rev. B}\ }\textbf {\bibinfo {volume}
  {81}},\ \bibinfo {pages} {041403} (\bibinfo {year} {2010})}\BibitemShut
  {NoStop}%
\bibitem [{\citenamefont {Zhou}\ \emph {et~al.}(2008)\citenamefont {Zhou},
  \citenamefont {Siegel}, \citenamefont {Fedorov},\ and\ \citenamefont
  {Lanzara}}]{zhou:2008}%
  \BibitemOpen
  \bibfield  {author} {\bibinfo {author} {\bibfnamefont {S.~Y.}\ \bibnamefont
  {Zhou}}, \bibinfo {author} {\bibfnamefont {D.~A.}\ \bibnamefont {Siegel}},
  \bibinfo {author} {\bibfnamefont {A.~V.}\ \bibnamefont {Fedorov}}, \ and\
  \bibinfo {author} {\bibfnamefont {A.}~\bibnamefont {Lanzara}},\ }\href@noop
  {} {\bibfield  {journal} {\bibinfo  {journal} {Phys. Rev. B}\ }\textbf
  {\bibinfo {volume} {78}},\ \bibinfo {pages} {193404} (\bibinfo {year}
  {2008})}\BibitemShut {NoStop}%
\bibitem [{\citenamefont {Carbotte}\ \emph {et~al.}(2011)\citenamefont
  {Carbotte}, \citenamefont {Timusk},\ and\ \citenamefont
  {Hwang}}]{carbotte:2011}%
  \BibitemOpen
  \bibfield  {author} {\bibinfo {author} {\bibfnamefont {J.~P.}\ \bibnamefont
  {Carbotte}}, \bibinfo {author} {\bibfnamefont {T.}~\bibnamefont {Timusk}}, \
  and\ \bibinfo {author} {\bibfnamefont {J.}~\bibnamefont {Hwang}},\
  }\href@noop {} {\bibfield  {journal} {\bibinfo  {journal} {Reports on
  Progress in Physics}\ }\textbf {\bibinfo {volume} {74}},\ \bibinfo {pages}
  {066501} (\bibinfo {year} {2011})}\BibitemShut {NoStop}%
\bibitem [{\citenamefont {Schachinger}\ and\ \citenamefont
  {Carbotte}(2000)}]{schachinger:2000}%
  \BibitemOpen
  \bibfield  {author} {\bibinfo {author} {\bibfnamefont {E.}~\bibnamefont
  {Schachinger}}\ and\ \bibinfo {author} {\bibfnamefont {J.~P.}\ \bibnamefont
  {Carbotte}},\ }\href@noop {} {\bibfield  {journal} {\bibinfo  {journal}
  {Phys. Rev. B}\ }\textbf {\bibinfo {volume} {62}},\ \bibinfo {pages} {9054}
  (\bibinfo {year} {2000})}\BibitemShut {NoStop}%
\bibitem [{\citenamefont {Schachinger}\ \emph {et~al.}(2003)\citenamefont
  {Schachinger}, \citenamefont {Tu},\ and\ \citenamefont
  {Carbotte}}]{schachinger:2003}%
  \BibitemOpen
  \bibfield  {author} {\bibinfo {author} {\bibfnamefont {E.}~\bibnamefont
  {Schachinger}}, \bibinfo {author} {\bibfnamefont {J.~J.}\ \bibnamefont {Tu}},
  \ and\ \bibinfo {author} {\bibfnamefont {J.~P.}\ \bibnamefont {Carbotte}},\
  }\href@noop {} {\bibfield  {journal} {\bibinfo  {journal} {Phys. Rev. B}\
  }\textbf {\bibinfo {volume} {67}},\ \bibinfo {pages} {214508} (\bibinfo
  {year} {2003})}\BibitemShut {NoStop}%
\bibitem [{\citenamefont {Li}\ \emph {et~al.}(2009)\citenamefont {Li},
  \citenamefont {Luican},\ and\ \citenamefont {Andrei}}]{li:2009}%
  \BibitemOpen
  \bibfield  {author} {\bibinfo {author} {\bibfnamefont {G.}~\bibnamefont
  {Li}}, \bibinfo {author} {\bibfnamefont {A.}~\bibnamefont {Luican}}, \ and\
  \bibinfo {author} {\bibfnamefont {E.~Y.}\ \bibnamefont {Andrei}},\
  }\href@noop {} {\bibfield  {journal} {\bibinfo  {journal} {Phys. Rev. Lett.}\
  }\textbf {\bibinfo {volume} {102}},\ \bibinfo {pages} {176804} (\bibinfo
  {year} {2009})}\BibitemShut {NoStop}%
\bibitem [{\citenamefont {Miller}\ \emph {et~al.}(2009)\citenamefont {Miller},
  \citenamefont {Kubista}, \citenamefont {Rutter}, \citenamefont {Ruan},
  \citenamefont {de~Heer}, \citenamefont {First},\ and\ \citenamefont
  {Stroscio}}]{miller:2009}%
  \BibitemOpen
  \bibfield  {author} {\bibinfo {author} {\bibfnamefont {D.~L.}\ \bibnamefont
  {Miller}}, \bibinfo {author} {\bibfnamefont {K.~D.}\ \bibnamefont {Kubista}},
  \bibinfo {author} {\bibfnamefont {G.~M.}\ \bibnamefont {Rutter}}, \bibinfo
  {author} {\bibfnamefont {M.}~\bibnamefont {Ruan}}, \bibinfo {author}
  {\bibfnamefont {W.~A.}\ \bibnamefont {de~Heer}}, \bibinfo {author}
  {\bibfnamefont {P.~N.}\ \bibnamefont {First}}, \ and\ \bibinfo {author}
  {\bibfnamefont {J.~A.}\ \bibnamefont {Stroscio}},\ }\href@noop {} {\bibfield
  {journal} {\bibinfo  {journal} {Science}\ }\textbf {\bibinfo {volume}
  {324}},\ \bibinfo {pages} {924} (\bibinfo {year} {2009})}\BibitemShut
  {NoStop}%
\bibitem [{\citenamefont {Brar}\ \emph {et~al.}(2010)\citenamefont {Brar},
  \citenamefont {Wickenburg}, \citenamefont {Panlasigui}, \citenamefont {Park},
  \citenamefont {Wehling}, \citenamefont {Zhang}, \citenamefont {Decker},
  \citenamefont {\c{C}a\u{g}lar Girit}, \citenamefont {Balatsky}, \citenamefont
  {Louie}, \citenamefont {Zettl},\ and\ \citenamefont {Crommie}}]{brar:2010}%
  \BibitemOpen
  \bibfield  {author} {\bibinfo {author} {\bibfnamefont {V.~W.}\ \bibnamefont
  {Brar}}, \bibinfo {author} {\bibfnamefont {S.}~\bibnamefont {Wickenburg}},
  \bibinfo {author} {\bibfnamefont {M.}~\bibnamefont {Panlasigui}}, \bibinfo
  {author} {\bibfnamefont {C.-H.}\ \bibnamefont {Park}}, \bibinfo {author}
  {\bibfnamefont {T.~O.}\ \bibnamefont {Wehling}}, \bibinfo {author}
  {\bibfnamefont {Y.}~\bibnamefont {Zhang}}, \bibinfo {author} {\bibfnamefont
  {R.}~\bibnamefont {Decker}}, \bibinfo {author} {\bibnamefont {\c{C}a\u{g}lar
  Girit}}, \bibinfo {author} {\bibfnamefont {A.~V.}\ \bibnamefont {Balatsky}},
  \bibinfo {author} {\bibfnamefont {S.~G.}\ \bibnamefont {Louie}}, \bibinfo
  {author} {\bibfnamefont {A.}~\bibnamefont {Zettl}}, \ and\ \bibinfo {author}
  {\bibfnamefont {M.~F.}\ \bibnamefont {Crommie}},\ }\href@noop {} {\bibfield
  {journal} {\bibinfo  {journal} {Phys. Rev. Lett.}\ }\textbf {\bibinfo
  {volume} {104}},\ \bibinfo {pages} {036805} (\bibinfo {year}
  {2010})}\BibitemShut {NoStop}%
\bibitem [{\citenamefont {Pound}\ \emph {et~al.}(2011)\citenamefont {Pound},
  \citenamefont {Carbotte},\ and\ \citenamefont {Nicol}}]{pound:2011}%
  \BibitemOpen
  \bibfield  {author} {\bibinfo {author} {\bibfnamefont {A.}~\bibnamefont
  {Pound}}, \bibinfo {author} {\bibfnamefont {J.~P.}\ \bibnamefont {Carbotte}},
  \ and\ \bibinfo {author} {\bibfnamefont {E.~J.}\ \bibnamefont {Nicol}},\
  }\href@noop {} {\bibfield  {journal} {\bibinfo  {journal} {Europhysics
  Letters}\ }\textbf {\bibinfo {volume} {94}},\ \bibinfo {pages} {57006}
  (\bibinfo {year} {2011})}\BibitemShut {NoStop}%
\bibitem [{\citenamefont {Nicol}\ and\ \citenamefont
  {Carbotte}(2009)}]{nicol:2009}%
  \BibitemOpen
  \bibfield  {author} {\bibinfo {author} {\bibfnamefont {E.~J.}\ \bibnamefont
  {Nicol}}\ and\ \bibinfo {author} {\bibfnamefont {J.~P.}\ \bibnamefont
  {Carbotte}},\ }\href@noop {} {\bibfield  {journal} {\bibinfo  {journal}
  {Phys. Rev. B}\ }\textbf {\bibinfo {volume} {80}},\ \bibinfo {pages}
  {081415(R)} (\bibinfo {year} {2009})}\BibitemShut {NoStop}%
\bibitem [{\citenamefont {Mitrovic}\ and\ \citenamefont
  {Carbotte}(1983{\natexlab{a}})}]{mitrovic:1983}%
  \BibitemOpen
  \bibfield  {author} {\bibinfo {author} {\bibfnamefont {B.}~\bibnamefont
  {Mitrovic}}\ and\ \bibinfo {author} {\bibfnamefont {J.~P.}\ \bibnamefont
  {Carbotte}},\ }\href@noop {} {\bibfield  {journal} {\bibinfo  {journal} {Can.
  J. Phys.}\ }\textbf {\bibinfo {volume} {61}},\ \bibinfo {pages} {758}
  (\bibinfo {year} {1983}{\natexlab{a}})}\BibitemShut {NoStop}%
\bibitem [{\citenamefont {Mitrovic}\ and\ \citenamefont
  {Carbotte}(1983{\natexlab{b}})}]{mitrovic:1983:2}%
  \BibitemOpen
  \bibfield  {author} {\bibinfo {author} {\bibfnamefont {B.}~\bibnamefont
  {Mitrovic}}\ and\ \bibinfo {author} {\bibfnamefont {J.~P.}\ \bibnamefont
  {Carbotte}},\ }\href@noop {} {\bibfield  {journal} {\bibinfo  {journal} {Can.
  J. Phys.}\ }\textbf {\bibinfo {volume} {61}},\ \bibinfo {pages} {784}
  (\bibinfo {year} {1983}{\natexlab{b}})}\BibitemShut {NoStop}%
\bibitem [{\citenamefont {Bostwick}\ \emph {et~al.}(2010)\citenamefont
  {Bostwick}, \citenamefont {Speck}, \citenamefont {Seyller}, \citenamefont
  {Horn}, \citenamefont {Polini}, \citenamefont {Asgari}, \citenamefont
  {Macdonald},\ and\ \citenamefont {Rotenberg}}]{bostwick:2010}%
  \BibitemOpen
  \bibfield  {author} {\bibinfo {author} {\bibfnamefont {A.}~\bibnamefont
  {Bostwick}}, \bibinfo {author} {\bibfnamefont {F.}~\bibnamefont {Speck}},
  \bibinfo {author} {\bibfnamefont {T.}~\bibnamefont {Seyller}}, \bibinfo
  {author} {\bibfnamefont {K.}~\bibnamefont {Horn}}, \bibinfo {author}
  {\bibfnamefont {M.}~\bibnamefont {Polini}}, \bibinfo {author} {\bibfnamefont
  {R.}~\bibnamefont {Asgari}}, \bibinfo {author} {\bibfnamefont {A.~H.}\
  \bibnamefont {Macdonald}}, \ and\ \bibinfo {author} {\bibfnamefont
  {E.}~\bibnamefont {Rotenberg}},\ }\href@noop {} {\bibfield  {journal}
  {\bibinfo  {journal} {Science}\ }\textbf {\bibinfo {volume} {328}},\ \bibinfo
  {pages} {999} (\bibinfo {year} {2010})}\BibitemShut {NoStop}%
\bibitem [{\citenamefont {Carbotte}\ \emph {et~al.}(1995)\citenamefont
  {Carbotte}, \citenamefont {Jiang}, \citenamefont {Basov},\ and\ \citenamefont
  {Timusk}}]{carbotte:1995}%
  \BibitemOpen
  \bibfield  {author} {\bibinfo {author} {\bibfnamefont {J.~P.}\ \bibnamefont
  {Carbotte}}, \bibinfo {author} {\bibfnamefont {C.}~\bibnamefont {Jiang}},
  \bibinfo {author} {\bibfnamefont {D.~N.}\ \bibnamefont {Basov}}, \ and\
  \bibinfo {author} {\bibfnamefont {T.}~\bibnamefont {Timusk}},\ }\href@noop {}
  {\bibfield  {journal} {\bibinfo  {journal} {Phys. Rev. B}\ }\textbf {\bibinfo
  {volume} {51}},\ \bibinfo {pages} {11798} (\bibinfo {year}
  {1995})}\BibitemShut {NoStop}%
\bibitem [{\citenamefont {Nicol}\ \emph {et~al.}(1991)\citenamefont {Nicol},
  \citenamefont {Carbotte},\ and\ \citenamefont {Timusk}}]{nicol:1991}%
  \BibitemOpen
  \bibfield  {author} {\bibinfo {author} {\bibfnamefont {E.~J.}\ \bibnamefont
  {Nicol}}, \bibinfo {author} {\bibfnamefont {J.~P.}\ \bibnamefont {Carbotte}},
  \ and\ \bibinfo {author} {\bibfnamefont {T.}~\bibnamefont {Timusk}},\
  }\href@noop {} {\bibfield  {journal} {\bibinfo  {journal} {Phys. Rev. B}\
  }\textbf {\bibinfo {volume} {43}},\ \bibinfo {pages} {473} (\bibinfo {year}
  {1991})}\BibitemShut {NoStop}%
\bibitem [{\citenamefont {Nicol}\ and\ \citenamefont
  {Carbotte}(1991)}]{nicol:1991:2}%
  \BibitemOpen
  \bibfield  {author} {\bibinfo {author} {\bibfnamefont {E.~J.}\ \bibnamefont
  {Nicol}}\ and\ \bibinfo {author} {\bibfnamefont {J.~P.}\ \bibnamefont
  {Carbotte}},\ }\href@noop {} {\bibfield  {journal} {\bibinfo  {journal}
  {Phys. Rev. B}\ }\textbf {\bibinfo {volume} {44}},\ \bibinfo {pages} {7741}
  (\bibinfo {year} {1991})}\BibitemShut {NoStop}%
\bibitem [{\citenamefont {Schachinger}\ \emph {et~al.}(1997)\citenamefont
  {Schachinger}, \citenamefont {Carbotte},\ and\ \citenamefont
  {Marsiglio}}]{schachinger:1997}%
  \BibitemOpen
  \bibfield  {author} {\bibinfo {author} {\bibfnamefont {E.}~\bibnamefont
  {Schachinger}}, \bibinfo {author} {\bibfnamefont {J.~P.}\ \bibnamefont
  {Carbotte}}, \ and\ \bibinfo {author} {\bibfnamefont {F.}~\bibnamefont
  {Marsiglio}},\ }\href@noop {} {\bibfield  {journal} {\bibinfo  {journal}
  {Phys. Rev. B}\ }\textbf {\bibinfo {volume} {56}},\ \bibinfo {pages} {2738}
  (\bibinfo {year} {1997})}\BibitemShut {NoStop}%
\bibitem [{\citenamefont {Carbotte}\ \emph {et~al.}(1986)\citenamefont
  {Carbotte}, \citenamefont {Marsiglio},\ and\ \citenamefont
  {Mitrovic}}]{carbotte:1986}%
  \BibitemOpen
  \bibfield  {author} {\bibinfo {author} {\bibfnamefont {J.~P.}\ \bibnamefont
  {Carbotte}}, \bibinfo {author} {\bibfnamefont {F.}~\bibnamefont {Marsiglio}},
  \ and\ \bibinfo {author} {\bibfnamefont {B.}~\bibnamefont {Mitrovic}},\
  }\href@noop {} {\bibfield  {journal} {\bibinfo  {journal} {Phys. Rev. B}\
  }\textbf {\bibinfo {volume} {33}},\ \bibinfo {pages} {6135} (\bibinfo {year}
  {1986})}\BibitemShut {NoStop}%
\bibitem [{\citenamefont {Marsiglio}\ \emph {et~al.}(1992)\citenamefont
  {Marsiglio}, \citenamefont {Akis},\ and\ \citenamefont
  {Carbotte}}]{marsiglio:1992}%
  \BibitemOpen
  \bibfield  {author} {\bibinfo {author} {\bibfnamefont {F.}~\bibnamefont
  {Marsiglio}}, \bibinfo {author} {\bibfnamefont {R.}~\bibnamefont {Akis}}, \
  and\ \bibinfo {author} {\bibfnamefont {J.~P.}\ \bibnamefont {Carbotte}},\
  }\href@noop {} {\bibfield  {journal} {\bibinfo  {journal} {Phys. Rev. B}\
  }\textbf {\bibinfo {volume} {45}},\ \bibinfo {pages} {9865} (\bibinfo {year}
  {1992})}\BibitemShut {NoStop}%
\bibitem [{\citenamefont {Mitrovic}\ \emph {et~al.}(1980)\citenamefont
  {Mitrovic}, \citenamefont {Leavens},\ and\ \citenamefont
  {Carbotte}}]{mitrovic:1980}%
  \BibitemOpen
  \bibfield  {author} {\bibinfo {author} {\bibfnamefont {B.}~\bibnamefont
  {Mitrovic}}, \bibinfo {author} {\bibfnamefont {C.~R.}\ \bibnamefont
  {Leavens}}, \ and\ \bibinfo {author} {\bibfnamefont {J.~P.}\ \bibnamefont
  {Carbotte}},\ }\href@noop {} {\bibfield  {journal} {\bibinfo  {journal}
  {Phys. Rev. B}\ }\textbf {\bibinfo {volume} {21}},\ \bibinfo {pages} {5048}
  (\bibinfo {year} {1980})}\BibitemShut {NoStop}%
\bibitem [{\citenamefont {Schachinger}\ \emph {et~al.}(1990)\citenamefont
  {Schachinger}, \citenamefont {Greeson},\ and\ \citenamefont
  {Carbotte}}]{schachinger:1990}%
  \BibitemOpen
  \bibfield  {author} {\bibinfo {author} {\bibfnamefont {E.}~\bibnamefont
  {Schachinger}}, \bibinfo {author} {\bibfnamefont {M.~G.}\ \bibnamefont
  {Greeson}}, \ and\ \bibinfo {author} {\bibfnamefont {J.~P.}\ \bibnamefont
  {Carbotte}},\ }\href@noop {} {\bibfield  {journal} {\bibinfo  {journal}
  {Phys. Rev. B}\ }\textbf {\bibinfo {volume} {42}},\ \bibinfo {pages} {406}
  (\bibinfo {year} {1990})}\BibitemShut {NoStop}%
\bibitem [{\citenamefont {Arberg}\ \emph {et~al.}(1993)\citenamefont {Arberg},
  \citenamefont {Mansor},\ and\ \citenamefont {Carbotte}}]{arberg:1993}%
  \BibitemOpen
  \bibfield  {author} {\bibinfo {author} {\bibfnamefont {P.}~\bibnamefont
  {Arberg}}, \bibinfo {author} {\bibfnamefont {M.}~\bibnamefont {Mansor}}, \
  and\ \bibinfo {author} {\bibfnamefont {J.~P.}\ \bibnamefont {Carbotte}},\
  }\href@noop {} {\bibfield  {journal} {\bibinfo  {journal} {Solid State
  Communications}\ }\textbf {\bibinfo {volume} {86}},\ \bibinfo {pages} {671}
  (\bibinfo {year} {1993})}\BibitemShut {NoStop}%
\bibitem [{\citenamefont {Branch}\ and\ \citenamefont
  {Carbotte}(1995)}]{branch:1995}%
  \BibitemOpen
  \bibfield  {author} {\bibinfo {author} {\bibfnamefont {D.}~\bibnamefont
  {Branch}}\ and\ \bibinfo {author} {\bibfnamefont {J.~P.}\ \bibnamefont
  {Carbotte}},\ }\href@noop {} {\bibfield  {journal} {\bibinfo  {journal}
  {Phys. Rev. B}\ }\textbf {\bibinfo {volume} {52}},\ \bibinfo {pages} {603}
  (\bibinfo {year} {1995})}\BibitemShut {NoStop}%
\bibitem [{\citenamefont {Leung}\ \emph {et~al.}(1976)\citenamefont {Leung},
  \citenamefont {Carbotte}, \citenamefont {Taylor},\ and\ \citenamefont
  {Leavens}}]{leung:1976}%
  \BibitemOpen
  \bibfield  {author} {\bibinfo {author} {\bibfnamefont {H.~K.}\ \bibnamefont
  {Leung}}, \bibinfo {author} {\bibfnamefont {J.~P.}\ \bibnamefont {Carbotte}},
  \bibinfo {author} {\bibfnamefont {D.~W.}\ \bibnamefont {Taylor}}, \ and\
  \bibinfo {author} {\bibfnamefont {C.~R.}\ \bibnamefont {Leavens}},\
  }\href@noop {} {\bibfield  {journal} {\bibinfo  {journal} {Canadian Journal
  of Physics}\ }\textbf {\bibinfo {volume} {54}},\ \bibinfo {pages} {1585}
  (\bibinfo {year} {1976})}\BibitemShut {NoStop}%
\bibitem [{\citenamefont {O'Donovan}\ and\ \citenamefont
  {Carbotte}(1995)}]{odonovan:1995}%
  \BibitemOpen
  \bibfield  {author} {\bibinfo {author} {\bibfnamefont {C.}~\bibnamefont
  {O'Donovan}}\ and\ \bibinfo {author} {\bibfnamefont {J.~P.}\ \bibnamefont
  {Carbotte}},\ }\href@noop {} {\bibfield  {journal} {\bibinfo  {journal}
  {Phys. Rev. B}\ }\textbf {\bibinfo {volume} {52}},\ \bibinfo {pages} {16208}
  (\bibinfo {year} {1995})}\BibitemShut {NoStop}%
\bibitem [{\citenamefont {Park}\ \emph {et~al.}(2007)\citenamefont {Park},
  \citenamefont {Giustino}, \citenamefont {Cohen},\ and\ \citenamefont
  {Louie}}]{park:2007}%
  \BibitemOpen
  \bibfield  {author} {\bibinfo {author} {\bibfnamefont {C.-H.}\ \bibnamefont
  {Park}}, \bibinfo {author} {\bibfnamefont {F.}~\bibnamefont {Giustino}},
  \bibinfo {author} {\bibfnamefont {M.~L.}\ \bibnamefont {Cohen}}, \ and\
  \bibinfo {author} {\bibfnamefont {S.~G.}\ \bibnamefont {Louie}},\ }\href@noop
  {} {\bibfield  {journal} {\bibinfo  {journal} {Phys. Rev. Lett.}\ }\textbf
  {\bibinfo {volume} {99}},\ \bibinfo {pages} {086804} (\bibinfo {year}
  {2007})}\BibitemShut {NoStop}%
\bibitem [{\citenamefont {Park}\ \emph {et~al.}(2008)\citenamefont {Park},
  \citenamefont {Giustino}, \citenamefont {Spataru}, \citenamefont {Cohen},\
  and\ \citenamefont {Louie}}]{park:2008}%
  \BibitemOpen
  \bibfield  {author} {\bibinfo {author} {\bibfnamefont {C.-H.}\ \bibnamefont
  {Park}}, \bibinfo {author} {\bibfnamefont {F.}~\bibnamefont {Giustino}},
  \bibinfo {author} {\bibfnamefont {C.~D.}\ \bibnamefont {Spataru}}, \bibinfo
  {author} {\bibfnamefont {M.~L.}\ \bibnamefont {Cohen}}, \ and\ \bibinfo
  {author} {\bibfnamefont {S.~G.}\ \bibnamefont {Louie}},\ }\href@noop {}
  {\bibfield  {journal} {\bibinfo  {journal} {Nano Lett.}\ }\textbf {\bibinfo
  {volume} {8}},\ \bibinfo {pages} {4229} (\bibinfo {year} {2008})}\BibitemShut
  {NoStop}%
\bibitem [{\citenamefont {Park}\ \emph {et~al.}(2009)\citenamefont {Park},
  \citenamefont {Giustino}, \citenamefont {Spataru}, \citenamefont {Cohen},\
  and\ \citenamefont {Louie}}]{park:2009}%
  \BibitemOpen
  \bibfield  {author} {\bibinfo {author} {\bibfnamefont {C.-H.}\ \bibnamefont
  {Park}}, \bibinfo {author} {\bibfnamefont {F.}~\bibnamefont {Giustino}},
  \bibinfo {author} {\bibfnamefont {C.~D.}\ \bibnamefont {Spataru}}, \bibinfo
  {author} {\bibfnamefont {M.~L.}\ \bibnamefont {Cohen}}, \ and\ \bibinfo
  {author} {\bibfnamefont {S.~G.}\ \bibnamefont {Louie}},\ }\href@noop {}
  {\bibfield  {journal} {\bibinfo  {journal} {Phys. Rev. Lett.}\ }\textbf
  {\bibinfo {volume} {102}},\ \bibinfo {pages} {076803} (\bibinfo {year}
  {2009})}\BibitemShut {NoStop}%
\bibitem [{\citenamefont {Li}\ \emph {et~al.}(2008)\citenamefont {Li},
  \citenamefont {Henriksen}, \citenamefont {Jiang}, \citenamefont {Hao},
  \citenamefont {Martin}, \citenamefont {Kim}, \citenamefont {Stormer},\ and\
  \citenamefont {Basov}}]{li:2008}%
  \BibitemOpen
  \bibfield  {author} {\bibinfo {author} {\bibfnamefont {Z.}~\bibnamefont
  {Li}}, \bibinfo {author} {\bibfnamefont {E.~A.}\ \bibnamefont {Henriksen}},
  \bibinfo {author} {\bibfnamefont {Z.}~\bibnamefont {Jiang}}, \bibinfo
  {author} {\bibfnamefont {Z.}~\bibnamefont {Hao}}, \bibinfo {author}
  {\bibfnamefont {M.~C.}\ \bibnamefont {Martin}}, \bibinfo {author}
  {\bibfnamefont {P.}~\bibnamefont {Kim}}, \bibinfo {author} {\bibfnamefont
  {H.~L.}\ \bibnamefont {Stormer}}, \ and\ \bibinfo {author} {\bibfnamefont
  {D.~N.}\ \bibnamefont {Basov}},\ }\href@noop {} {\bibfield  {journal}
  {\bibinfo  {journal} {Nat. Phys.}\ }\textbf {\bibinfo {volume} {4}},\
  \bibinfo {pages} {532} (\bibinfo {year} {2008})}\BibitemShut {NoStop}%
\bibitem [{\citenamefont {Wang}\ \emph {et~al.}(2008)\citenamefont {Wang},
  \citenamefont {Zhang}, \citenamefont {Tian}, \citenamefont {Girit},
  \citenamefont {Zettl}, \citenamefont {Crommie},\ and\ \citenamefont
  {Shen}}]{wang:2008}%
  \BibitemOpen
  \bibfield  {author} {\bibinfo {author} {\bibfnamefont {F.}~\bibnamefont
  {Wang}}, \bibinfo {author} {\bibfnamefont {Y.}~\bibnamefont {Zhang}},
  \bibinfo {author} {\bibfnamefont {C.}~\bibnamefont {Tian}}, \bibinfo {author}
  {\bibfnamefont {C.}~\bibnamefont {Girit}}, \bibinfo {author} {\bibfnamefont
  {A.}~\bibnamefont {Zettl}}, \bibinfo {author} {\bibfnamefont
  {M.}~\bibnamefont {Crommie}}, \ and\ \bibinfo {author} {\bibfnamefont
  {Y.~R.}\ \bibnamefont {Shen}},\ }\href@noop {} {\bibfield  {journal}
  {\bibinfo  {journal} {Science}\ }\textbf {\bibinfo {volume} {320}},\ \bibinfo
  {pages} {206} (\bibinfo {year} {2008})}\BibitemShut {NoStop}%
\bibitem [{\citenamefont {Mak}\ \emph {et~al.}(2008)\citenamefont {Mak},
  \citenamefont {Sfeir}, \citenamefont {Wu}, \citenamefont {Lui}, \citenamefont
  {Misewich},\ and\ \citenamefont {Heinz}}]{mak:2008}%
  \BibitemOpen
  \bibfield  {author} {\bibinfo {author} {\bibfnamefont {K.~F.}\ \bibnamefont
  {Mak}}, \bibinfo {author} {\bibfnamefont {M.~Y.}\ \bibnamefont {Sfeir}},
  \bibinfo {author} {\bibfnamefont {Y.}~\bibnamefont {Wu}}, \bibinfo {author}
  {\bibfnamefont {C.~H.}\ \bibnamefont {Lui}}, \bibinfo {author} {\bibfnamefont
  {J.~A.}\ \bibnamefont {Misewich}}, \ and\ \bibinfo {author} {\bibfnamefont
  {T.~F.}\ \bibnamefont {Heinz}},\ }\href@noop {} {\bibfield  {journal}
  {\bibinfo  {journal} {Phys. Rev. Lett.}\ }\textbf {\bibinfo {volume} {101}},\
  \bibinfo {pages} {196405} (\bibinfo {year} {2008})}\BibitemShut {NoStop}%
\bibitem [{\citenamefont {Nair}\ \emph {et~al.}(2008)\citenamefont {Nair},
  \citenamefont {Blake}, \citenamefont {Grigorenko}, \citenamefont {Novoselov},
  \citenamefont {Booth}, \citenamefont {Stauber}, \citenamefont {Peres},\ and\
  \citenamefont {Geim}}]{nair:2008}%
  \BibitemOpen
  \bibfield  {author} {\bibinfo {author} {\bibfnamefont {R.~R.}\ \bibnamefont
  {Nair}}, \bibinfo {author} {\bibfnamefont {P.}~\bibnamefont {Blake}},
  \bibinfo {author} {\bibfnamefont {A.~N.}\ \bibnamefont {Grigorenko}},
  \bibinfo {author} {\bibfnamefont {K.~S.}\ \bibnamefont {Novoselov}}, \bibinfo
  {author} {\bibfnamefont {T.~J.}\ \bibnamefont {Booth}}, \bibinfo {author}
  {\bibfnamefont {T.}~\bibnamefont {Stauber}}, \bibinfo {author} {\bibfnamefont
  {N.~M.~R.}\ \bibnamefont {Peres}}, \ and\ \bibinfo {author} {\bibfnamefont
  {A.~K.}\ \bibnamefont {Geim}},\ }\href@noop {} {\bibfield  {journal}
  {\bibinfo  {journal} {Science}\ }\textbf {\bibinfo {volume} {320}},\ \bibinfo
  {pages} {1308} (\bibinfo {year} {2008})}\BibitemShut {NoStop}%
\bibitem [{\citenamefont {Peres}\ \emph {et~al.}(2008)\citenamefont {Peres},
  \citenamefont {Stauber},\ and\ \citenamefont {{Castro Neto}}}]{peres:2008}%
  \BibitemOpen
  \bibfield  {author} {\bibinfo {author} {\bibfnamefont {N.~M.~R.}\
  \bibnamefont {Peres}}, \bibinfo {author} {\bibfnamefont {T.}~\bibnamefont
  {Stauber}}, \ and\ \bibinfo {author} {\bibfnamefont {A.~H.}\ \bibnamefont
  {{Castro Neto}}},\ }\href@noop {} {\bibfield  {journal} {\bibinfo  {journal}
  {EPL}\ }\textbf {\bibinfo {volume} {84}},\ \bibinfo {pages} {38002} (\bibinfo
  {year} {2008})}\BibitemShut {NoStop}%
\bibitem [{\citenamefont {Carbotte}\ \emph {et~al.}(2010)\citenamefont
  {Carbotte}, \citenamefont {Nicol},\ and\ \citenamefont
  {Sharapov}}]{carbotte:2010}%
  \BibitemOpen
  \bibfield  {author} {\bibinfo {author} {\bibfnamefont {J.~P.}\ \bibnamefont
  {Carbotte}}, \bibinfo {author} {\bibfnamefont {E.~J.}\ \bibnamefont {Nicol}},
  \ and\ \bibinfo {author} {\bibfnamefont {S.~G.}\ \bibnamefont {Sharapov}},\
  }\href@noop {} {\bibfield  {journal} {\bibinfo  {journal} {Phys. Rev. B}\
  }\textbf {\bibinfo {volume} {81}},\ \bibinfo {pages} {045419} (\bibinfo
  {year} {2010})}\BibitemShut {NoStop}%
\bibitem [{\citenamefont {Grushin}\ \emph {et~al.}(2009)\citenamefont
  {Grushin}, \citenamefont {Valenzuela},\ and\ \citenamefont
  {Vozmediano}}]{grushin:2009}%
  \BibitemOpen
  \bibfield  {author} {\bibinfo {author} {\bibfnamefont {A.~G.}\ \bibnamefont
  {Grushin}}, \bibinfo {author} {\bibfnamefont {B.}~\bibnamefont {Valenzuela}},
  \ and\ \bibinfo {author} {\bibfnamefont {M.~A.~H.}\ \bibnamefont
  {Vozmediano}},\ }\href@noop {} {\bibfield  {journal} {\bibinfo  {journal}
  {Phys. Rev. B}\ }\textbf {\bibinfo {volume} {80}},\ \bibinfo {pages} {155417}
  (\bibinfo {year} {2009})}\BibitemShut {NoStop}%
\bibitem [{\citenamefont {Nicol}\ and\ \citenamefont
  {Carbotte}(2008)}]{nicol:2008}%
  \BibitemOpen
  \bibfield  {author} {\bibinfo {author} {\bibfnamefont {E.~J.}\ \bibnamefont
  {Nicol}}\ and\ \bibinfo {author} {\bibfnamefont {J.~P.}\ \bibnamefont
  {Carbotte}},\ }\href@noop {} {\bibfield  {journal} {\bibinfo  {journal}
  {Phys. Rev. B}\ }\textbf {\bibinfo {volume} {77}},\ \bibinfo {pages} {155409}
  (\bibinfo {year} {2008})}\BibitemShut {NoStop}%
\bibitem [{\citenamefont {Kuzmenko}\ \emph {et~al.}(2009)\citenamefont
  {Kuzmenko}, \citenamefont {{van Heumen}}, \citenamefont {{van der Marel}},
  \citenamefont {Lerch}, \citenamefont {Blake}, \citenamefont {Novoselov},\
  and\ \citenamefont {Geim}}]{kuzmenko:2009}%
  \BibitemOpen
  \bibfield  {author} {\bibinfo {author} {\bibfnamefont {A.~B.}\ \bibnamefont
  {Kuzmenko}}, \bibinfo {author} {\bibfnamefont {E.}~\bibnamefont {{van
  Heumen}}}, \bibinfo {author} {\bibfnamefont {D.}~\bibnamefont {{van der
  Marel}}}, \bibinfo {author} {\bibfnamefont {P.}~\bibnamefont {Lerch}},
  \bibinfo {author} {\bibfnamefont {P.}~\bibnamefont {Blake}}, \bibinfo
  {author} {\bibfnamefont {K.~S.}\ \bibnamefont {Novoselov}}, \ and\ \bibinfo
  {author} {\bibfnamefont {A.~K.}\ \bibnamefont {Geim}},\ }\href@noop {}
  {\bibfield  {journal} {\bibinfo  {journal} {Phys. Rev. B}\ }\textbf {\bibinfo
  {volume} {79}},\ \bibinfo {pages} {115441} (\bibinfo {year}
  {2009})}\BibitemShut {NoStop}%
\bibitem [{\citenamefont {Henriksen}\ \emph {et~al.}(2008)\citenamefont
  {Henriksen}, \citenamefont {Jiang}, \citenamefont {Tung}, \citenamefont
  {Schwartz}, \citenamefont {Takita}, \citenamefont {Wang}, \citenamefont
  {Kim},\ and\ \citenamefont {Stormer}}]{henriksen:2008}%
  \BibitemOpen
  \bibfield  {author} {\bibinfo {author} {\bibfnamefont {E.~A.}\ \bibnamefont
  {Henriksen}}, \bibinfo {author} {\bibfnamefont {Z.}~\bibnamefont {Jiang}},
  \bibinfo {author} {\bibfnamefont {L.-C.}\ \bibnamefont {Tung}}, \bibinfo
  {author} {\bibfnamefont {M.~E.}\ \bibnamefont {Schwartz}}, \bibinfo {author}
  {\bibfnamefont {M.}~\bibnamefont {Takita}}, \bibinfo {author} {\bibfnamefont
  {Y.-J.}\ \bibnamefont {Wang}}, \bibinfo {author} {\bibfnamefont
  {P.}~\bibnamefont {Kim}}, \ and\ \bibinfo {author} {\bibfnamefont {H.~L.}\
  \bibnamefont {Stormer}},\ }\href@noop {} {\bibfield  {journal} {\bibinfo
  {journal} {Phys. Rev. Lett.}\ }\textbf {\bibinfo {volume} {100}},\ \bibinfo
  {pages} {087403} (\bibinfo {year} {2008})}\BibitemShut {NoStop}%
\bibitem [{\citenamefont {Wunsch}\ \emph {et~al.}(2006)\citenamefont {Wunsch},
  \citenamefont {Stauber}, \citenamefont {Sols},\ and\ \citenamefont
  {Guinea}}]{wunsch:2006}%
  \BibitemOpen
  \bibfield  {author} {\bibinfo {author} {\bibfnamefont {B.}~\bibnamefont
  {Wunsch}}, \bibinfo {author} {\bibfnamefont {T.}~\bibnamefont {Stauber}},
  \bibinfo {author} {\bibfnamefont {F.}~\bibnamefont {Sols}}, \ and\ \bibinfo
  {author} {\bibfnamefont {F.}~\bibnamefont {Guinea}},\ }\href@noop {}
  {\bibfield  {journal} {\bibinfo  {journal} {New Journal of Physics}\ }\textbf
  {\bibinfo {volume} {8}},\ \bibinfo {pages} {318} (\bibinfo {year}
  {2006})}\BibitemShut {NoStop}%
\bibitem [{\citenamefont {Polini}\ \emph {et~al.}(2008)\citenamefont {Polini},
  \citenamefont {Asgari}, \citenamefont {Borghi}, \citenamefont {Barlas},
  \citenamefont {Pereg-Barnea},\ and\ \citenamefont {MacDonald}}]{polini:2008}%
  \BibitemOpen
  \bibfield  {author} {\bibinfo {author} {\bibfnamefont {M.}~\bibnamefont
  {Polini}}, \bibinfo {author} {\bibfnamefont {R.}~\bibnamefont {Asgari}},
  \bibinfo {author} {\bibfnamefont {G.}~\bibnamefont {Borghi}}, \bibinfo
  {author} {\bibfnamefont {Y.}~\bibnamefont {Barlas}}, \bibinfo {author}
  {\bibfnamefont {T.}~\bibnamefont {Pereg-Barnea}}, \ and\ \bibinfo {author}
  {\bibfnamefont {A.~H.}\ \bibnamefont {MacDonald}},\ }\href@noop {} {\bibfield
   {journal} {\bibinfo  {journal} {Phys. Rev. B}\ }\textbf {\bibinfo {volume}
  {77}},\ \bibinfo {pages} {081411(R)} (\bibinfo {year} {2008})}\BibitemShut
  {NoStop}%
\bibitem [{\citenamefont {Hwang}\ and\ \citenamefont {{Das
  Sarma}}(2008)}]{hwang:2008}%
  \BibitemOpen
  \bibfield  {author} {\bibinfo {author} {\bibfnamefont {E.~H.}\ \bibnamefont
  {Hwang}}\ and\ \bibinfo {author} {\bibfnamefont {S.}~\bibnamefont {{Das
  Sarma}}},\ }\href@noop {} {\bibfield  {journal} {\bibinfo  {journal} {Phys.
  Rev. B}\ }\textbf {\bibinfo {volume} {77}},\ \bibinfo {pages} {081412(R)}
  (\bibinfo {year} {2008})}\BibitemShut {NoStop}%
\bibitem [{\citenamefont {Sensarma}\ \emph {et~al.}(2011)\citenamefont
  {Sensarma}, \citenamefont {Hwang},\ and\ \citenamefont {{Das
  Sarma}}}]{sensarma:2011}%
  \BibitemOpen
  \bibfield  {author} {\bibinfo {author} {\bibfnamefont {R.}~\bibnamefont
  {Sensarma}}, \bibinfo {author} {\bibfnamefont {E.~H.}\ \bibnamefont {Hwang}},
  \ and\ \bibinfo {author} {\bibfnamefont {S.}~\bibnamefont {{Das Sarma}}},\
  }\href@noop {} {\bibfield  {journal} {\bibinfo  {journal} {Phys. Rev. B}\
  }\textbf {\bibinfo {volume} {84}},\ \bibinfo {pages} {041408(R)} (\bibinfo
  {year} {2011})}\BibitemShut {NoStop}%
\bibitem [{\citenamefont {Barlas}\ \emph {et~al.}(2007)\citenamefont {Barlas},
  \citenamefont {Pereg-Barnea}, \citenamefont {Polini}, \citenamefont
  {Asgari},\ and\ \citenamefont {MacDonald}}]{barlas:2007}%
  \BibitemOpen
  \bibfield  {author} {\bibinfo {author} {\bibfnamefont {Y.}~\bibnamefont
  {Barlas}}, \bibinfo {author} {\bibfnamefont {T.}~\bibnamefont
  {Pereg-Barnea}}, \bibinfo {author} {\bibfnamefont {M.}~\bibnamefont
  {Polini}}, \bibinfo {author} {\bibfnamefont {R.}~\bibnamefont {Asgari}}, \
  and\ \bibinfo {author} {\bibfnamefont {A.~H.}\ \bibnamefont {MacDonald}},\
  }\href@noop {} {\bibfield  {journal} {\bibinfo  {journal} {Phys. Rev. Lett.}\
  }\textbf {\bibinfo {volume} {98}},\ \bibinfo {pages} {236601} (\bibinfo
  {year} {2007})}\BibitemShut {NoStop}%
\bibitem [{\citenamefont {Hwang}\ and\ \citenamefont {{Das
  Sarma}}(2007)}]{ehhwang:2007}%
  \BibitemOpen
  \bibfield  {author} {\bibinfo {author} {\bibfnamefont {E.~H.}\ \bibnamefont
  {Hwang}}\ and\ \bibinfo {author} {\bibfnamefont {S.}~\bibnamefont {{Das
  Sarma}}},\ }\href@noop {} {\bibfield  {journal} {\bibinfo  {journal} {Phys.
  Rev. B}\ }\textbf {\bibinfo {volume} {75}},\ \bibinfo {pages} {205418}
  (\bibinfo {year} {2007})}\BibitemShut {NoStop}%
\bibitem [{\citenamefont {LeBlanc}\ \emph {et~al.}(2011)\citenamefont
  {LeBlanc}, \citenamefont {Carbotte},\ and\ \citenamefont
  {Nicol}}]{leblanc:2011}%
  \BibitemOpen
  \bibfield  {author} {\bibinfo {author} {\bibfnamefont {J.~P.~F.}\
  \bibnamefont {LeBlanc}}, \bibinfo {author} {\bibfnamefont {J.~P.}\
  \bibnamefont {Carbotte}}, \ and\ \bibinfo {author} {\bibfnamefont {E.~J.}\
  \bibnamefont {Nicol}},\ }\href@noop {} {\bibfield  {journal} {\bibinfo
  {journal} {Phys. Rev. B}\ }\textbf {\bibinfo {volume} {84}},\ \bibinfo
  {pages} {165448} (\bibinfo {year} {2011})}\BibitemShut {NoStop}%
\bibitem [{\citenamefont {Principi}\ \emph {et~al.}(2011)\citenamefont
  {Principi}, \citenamefont {Polini}, \citenamefont {Asgari},\ and\
  \citenamefont {MacDonald}}]{principi:2011}%
  \BibitemOpen
  \bibfield  {author} {\bibinfo {author} {\bibfnamefont {A.}~\bibnamefont
  {Principi}}, \bibinfo {author} {\bibfnamefont {M.}~\bibnamefont {Polini}},
  \bibinfo {author} {\bibfnamefont {R.}~\bibnamefont {Asgari}}, \ and\ \bibinfo
  {author} {\bibfnamefont {A.~H.}\ \bibnamefont {MacDonald}},\ }\href@noop {}
  {\bibfield  {journal} {\bibinfo  {journal} {arXiv:cond-mat}\ ,\ \bibinfo
  {pages} {1111.3822v1}} (\bibinfo {year} {2011})}\BibitemShut {NoStop}%
\bibitem [{\citenamefont {Walter}\ \emph {et~al.}(2011)\citenamefont {Walter},
  \citenamefont {Bostwick}, \citenamefont {Jeon}, \citenamefont {Speck},
  \citenamefont {Ostler}, \citenamefont {Syller}, \citenamefont {Moreschini},
  \citenamefont {Chang}, \citenamefont {Polini}, \citenamefont {Asgari},
  \citenamefont {MacDonald}, \citenamefont {Horn},\ and\ \citenamefont
  {Rotenberg}}]{walter:2011}%
  \BibitemOpen
  \bibfield  {author} {\bibinfo {author} {\bibfnamefont {A.~L.}\ \bibnamefont
  {Walter}}, \bibinfo {author} {\bibfnamefont {A.}~\bibnamefont {Bostwick}},
  \bibinfo {author} {\bibfnamefont {K.-J.}\ \bibnamefont {Jeon}}, \bibinfo
  {author} {\bibfnamefont {F.}~\bibnamefont {Speck}}, \bibinfo {author}
  {\bibfnamefont {M.}~\bibnamefont {Ostler}}, \bibinfo {author} {\bibfnamefont
  {T.}~\bibnamefont {Syller}}, \bibinfo {author} {\bibfnamefont
  {L.}~\bibnamefont {Moreschini}}, \bibinfo {author} {\bibfnamefont {Y.~J.}\
  \bibnamefont {Chang}}, \bibinfo {author} {\bibfnamefont {M.}~\bibnamefont
  {Polini}}, \bibinfo {author} {\bibfnamefont {R.}~\bibnamefont {Asgari}},
  \bibinfo {author} {\bibfnamefont {A.~H.}\ \bibnamefont {MacDonald}}, \bibinfo
  {author} {\bibfnamefont {K.}~\bibnamefont {Horn}}, \ and\ \bibinfo {author}
  {\bibfnamefont {E.}~\bibnamefont {Rotenberg}},\ }\href@noop {} {\bibfield
  {journal} {\bibinfo  {journal} {arXiv cond-mat}\ ,\ \bibinfo {pages}
  {1107.4398}} (\bibinfo {year} {2011})}\BibitemShut {NoStop}%
\bibitem [{\citenamefont {Zhang}\ \emph {et~al.}(2008)\citenamefont {Zhang},
  \citenamefont {Liu}, \citenamefont {Zhao}, \citenamefont {Liu}, \citenamefont
  {Meng}, \citenamefont {Dong}, \citenamefont {Lu}, \citenamefont {Wen},
  \citenamefont {Xu}, \citenamefont {Gu}, \citenamefont {Sasagawa},
  \citenamefont {Wang}, \citenamefont {Zhu}, \citenamefont {Zhang},
  \citenamefont {Zhou}, \citenamefont {Wang}, \citenamefont {Zhao},
  \citenamefont {Chen}, \citenamefont {Xu},\ and\ \citenamefont
  {Zhou}}]{zhang:2008}%
  \BibitemOpen
  \bibfield  {author} {\bibinfo {author} {\bibfnamefont {W.}~\bibnamefont
  {Zhang}}, \bibinfo {author} {\bibfnamefont {G.}~\bibnamefont {Liu}}, \bibinfo
  {author} {\bibfnamefont {L.}~\bibnamefont {Zhao}}, \bibinfo {author}
  {\bibfnamefont {H.}~\bibnamefont {Liu}}, \bibinfo {author} {\bibfnamefont
  {J.}~\bibnamefont {Meng}}, \bibinfo {author} {\bibfnamefont {X.}~\bibnamefont
  {Dong}}, \bibinfo {author} {\bibfnamefont {W.}~\bibnamefont {Lu}}, \bibinfo
  {author} {\bibfnamefont {J.~S.}\ \bibnamefont {Wen}}, \bibinfo {author}
  {\bibfnamefont {Z.~J.}\ \bibnamefont {Xu}}, \bibinfo {author} {\bibfnamefont
  {G.~D.}\ \bibnamefont {Gu}}, \bibinfo {author} {\bibfnamefont
  {T.}~\bibnamefont {Sasagawa}}, \bibinfo {author} {\bibfnamefont
  {G.}~\bibnamefont {Wang}}, \bibinfo {author} {\bibfnamefont {Y.}~\bibnamefont
  {Zhu}}, \bibinfo {author} {\bibfnamefont {H.}~\bibnamefont {Zhang}}, \bibinfo
  {author} {\bibfnamefont {Y.}~\bibnamefont {Zhou}}, \bibinfo {author}
  {\bibfnamefont {X.}~\bibnamefont {Wang}}, \bibinfo {author} {\bibfnamefont
  {Z.}~\bibnamefont {Zhao}}, \bibinfo {author} {\bibfnamefont {C.}~\bibnamefont
  {Chen}}, \bibinfo {author} {\bibfnamefont {Z.}~\bibnamefont {Xu}}, \ and\
  \bibinfo {author} {\bibfnamefont {X.~J.}\ \bibnamefont {Zhou}},\ }\href@noop
  {} {\bibfield  {journal} {\bibinfo  {journal} {Phys. Rev. Lett.}\ }\textbf
  {\bibinfo {volume} {100}},\ \bibinfo {pages} {107002} (\bibinfo {year}
  {2008})}\BibitemShut {NoStop}%
\bibitem [{\citenamefont {Schachinger}\ \emph {et~al.}(2006)\citenamefont
  {Schachinger}, \citenamefont {Neuber},\ and\ \citenamefont
  {Carbotte}}]{schachinger:2006}%
  \BibitemOpen
  \bibfield  {author} {\bibinfo {author} {\bibfnamefont {E.}~\bibnamefont
  {Schachinger}}, \bibinfo {author} {\bibfnamefont {D.}~\bibnamefont {Neuber}},
  \ and\ \bibinfo {author} {\bibfnamefont {J.~P.}\ \bibnamefont {Carbotte}},\
  }\href@noop {} {\bibfield  {journal} {\bibinfo  {journal} {Phys. Rev. B}\
  }\textbf {\bibinfo {volume} {73}},\ \bibinfo {pages} {184507} (\bibinfo
  {year} {2006})}\BibitemShut {NoStop}%
\bibitem [{\citenamefont {Schachinger}\ and\ \citenamefont
  {Carbotte}(2008)}]{schachinger:2008}%
  \BibitemOpen
  \bibfield  {author} {\bibinfo {author} {\bibfnamefont {E.}~\bibnamefont
  {Schachinger}}\ and\ \bibinfo {author} {\bibfnamefont {J.~P.}\ \bibnamefont
  {Carbotte}},\ }\href@noop {} {\bibfield  {journal} {\bibinfo  {journal}
  {Phys. Rev. B}\ }\textbf {\bibinfo {volume} {77}},\ \bibinfo {pages} {094524}
  (\bibinfo {year} {2008})}\BibitemShut {NoStop}%
\bibitem [{\citenamefont {Wu}\ \emph {et~al.}(2010)\citenamefont {Wu},
  \citenamefont {Bari\u{s}i\'{c}}, \citenamefont {Dressel}, \citenamefont
  {Cao}, \citenamefont {Xu}, \citenamefont {Schachinger},\ and\ \citenamefont
  {Carbotte}}]{wu:2010}%
  \BibitemOpen
  \bibfield  {author} {\bibinfo {author} {\bibfnamefont {D.}~\bibnamefont
  {Wu}}, \bibinfo {author} {\bibfnamefont {N.}~\bibnamefont {Bari\u{s}i\'{c}}},
  \bibinfo {author} {\bibfnamefont {M.}~\bibnamefont {Dressel}}, \bibinfo
  {author} {\bibfnamefont {G.~H.}\ \bibnamefont {Cao}}, \bibinfo {author}
  {\bibfnamefont {Z.-A.}\ \bibnamefont {Xu}}, \bibinfo {author} {\bibfnamefont
  {E.}~\bibnamefont {Schachinger}}, \ and\ \bibinfo {author} {\bibfnamefont
  {J.~P.}\ \bibnamefont {Carbotte}},\ }\href@noop {} {\bibfield  {journal}
  {\bibinfo  {journal} {Phys. Rev. B}\ }\textbf {\bibinfo {volume} {82}},\
  \bibinfo {pages} {144519} (\bibinfo {year} {2010})}\BibitemShut {NoStop}%
\bibitem [{\citenamefont {Bok}\ \emph {et~al.}(2010)\citenamefont {Bok},
  \citenamefont {Yun}, \citenamefont {Choi}, \citenamefont {Zhang},
  \citenamefont {Zhou},\ and\ \citenamefont {Varma}}]{bok:2010}%
  \BibitemOpen
  \bibfield  {author} {\bibinfo {author} {\bibfnamefont {J.~M.}\ \bibnamefont
  {Bok}}, \bibinfo {author} {\bibfnamefont {J.~H.}\ \bibnamefont {Yun}},
  \bibinfo {author} {\bibfnamefont {H.-Y.}\ \bibnamefont {Choi}}, \bibinfo
  {author} {\bibfnamefont {W.}~\bibnamefont {Zhang}}, \bibinfo {author}
  {\bibfnamefont {X.~J.}\ \bibnamefont {Zhou}}, \ and\ \bibinfo {author}
  {\bibfnamefont {C.~M.}\ \bibnamefont {Varma}},\ }\href@noop {} {\bibfield
  {journal} {\bibinfo  {journal} {Phys. Rev. B}\ }\textbf {\bibinfo {volume}
  {81}},\ \bibinfo {pages} {174516} (\bibinfo {year} {2010})}\BibitemShut
  {NoStop}%
\end{thebibliography}%

\end{document}